\documentclass[10pt]{article}

\usepackage{amsmath}

\usepackage{array}

\usepackage{appendix}

\usepackage{tocloft}                   

\usepackage{graphicx}

\usepackage{amsfonts}

\usepackage{amssymb}

\usepackage{mathrsfs}

\usepackage{yfonts}

\usepackage{euscript}

\usepackage{centernot}                 

\usepackage{ifsym}                     

\usepackage{upgreek}

\usepackage{mathtools}

\usepackage{color}

\usepackage{slantsc}
\usepackage{calligra}

\usepackage{bbold}          

\usepackage[T1]{fontenc}

\usepackage{epsf}

\usepackage{latexsym}

\usepackage{tipa}

\usepackage{makeidx}

\makeindex

\def\be{\begin{equation}}

\def\ee{\end{equation}}

\def\beq{\begin{equation}}

\def\eeq{\end{equation}}

\def\bea{\begin{eqnarray}}

\def\eea{\end{eqnarray}}

\def\ni{\noindent}

\def\!{\hspace{-1.6667em}}

\def\m{\mbox{ }}

\def\mma {\m , \m \m }

\def\!{\hspace{-1.6667em}}

\def\n{\noindent}

\def\u{\underline}

\def\w{\widetilde}

\def\s{\stackrel}

\def\es{\m = \m}

\def\:={\m := \m}

\def\=:{\m =: \m}

\def\peq{\m \mbox{`='} \m}

\def\bGamma{\mbox{\boldmath$\Gamma$}}               

\def\bT{\mbox{\bf T}}

\def\bg{\mbox{\bf g}}

\def\bsigma{\mbox{\boldmath$\sigma$}}                   %

\def\bGamma{\mbox{\boldmath$\Gamma$}}

\def\sa{\mbox{\scriptsize a}}

\def\si{\mbox{\scriptsize i}}

\def\sll{\mbox{\scriptsize l}}  

\def\sr{\mbox{\scriptsize r}}

\def\st{\mbox{\scriptsize t}}

\def\sT{\mbox{\scriptsize T}}

\def\sfk{\mbox{\sffamily{\scriptsize k}}}     

\def\bigtau{\mbox{\Large$\tau$}}

\def\Thomas{\, \mbox{\textcircled{$\rightarrow$}} \,}

\def\TwoWay{\, \mbox{\textcircled{$\leftrightarrow$}} \,}

\def\sumi2{\sum\mbox{}_{\mbox{}_{\mbox{\scriptsize $i$=1}}}^2}

\def\sumi3{\sum\mbox{}_{\mbox{}_{\mbox{\scriptsize $i$=1}}}^3}

\def\sumABcycles3{\sum\mbox{}_{\mbox{}_{\mbox{\scriptsize cycles $A,B$=1}}}^{3}}

\def\sumCDcycles3{\sum\mbox{}_{\mbox{}_{\mbox{\scriptsize cycles $C,D$=1}}}^{3}}

\def\sumj3{\sum\mbox{}_{\mbox{}_{\mbox{\scriptsize $j$=1}}}^3}

\def\sumk3{\sum\mbox{}_{\mbox{}_{\mbox{\scriptsize $k$=1}}}^3}

\def\prodiA1{\prod\mbox{}_{\mbox{}_{\mbox{\scriptsize $i$=1}}}^{A - 1}}

\def\d{\textrm{d}}                                                  

\def\pa{\partial}                                                   

\def\FrM{\mbox{$\mathfrak{M}$}}                                
\def\sFrM{\mbox{\scriptsize$\mathfrak{M}$}}                    

\def\FrN{\mbox{$\mathfrak{N}$}}                                
\def\lFrg{\mbox{\Large$\mathfrak{g}$}}                         
\def\FrH{\mbox{$\mathfrak{H}$}}                                
\def\FrN{\mathfrak{N}}                                         
                                                  
\def\Frg{\mbox{\normalsize $\mathfrak{g}$}}                    

\def\Frk{\mbox{\scriptsize $\mathfrak{K}$}}                    

\def\Frh{\mbox{$\mathfrak{h}$}}                                

\def\Hilb{\mbox{{\boldmath$\mathfrak{H}$}ilb}}                 

\def\scC{\mbox{\scriptsize ${\cal C}$}}                    

\def\Phase{\mbox{{\boldmath$\mathfrak{P}$}hase}}                     

\def\bFrR{\mbox{\boldmath$\mathfrak{R}$}}                            
\def\Rig-Phase{\bFrR\mbox{ig-}\Phase}                                
                                                                													   
\def\bFrM{\mbox{\boldmath${\mathfrak{M}}$}}                             

\def\Positive-Modespace{\mbox{{\boldmath$\mathfrak{M}$}odespace$^+$}}

\def\POSITIVE-MODESPACE{\mbox{{\boldmath$\mathfrak{M}$}ODESPACE$^+$}}

\def\K{Kucha\v{r} }

\def\Kin-Hilb{\mbox{{\boldmath$\mathfrak{K}$}in-\Hilb}}                     

\def\Mid-Hilb{\mbox{{\boldmath$\mathfrak{M}$}id-\Hilb}}                     

\def\Dyn-Hilb{\mbox{{\boldmath$\mathfrak{D}$}yn-\Hilb}}                     

\def\5Star{\mbox{\Large$\star$}}              

\begin{document}

\begin{titlepage}

\begin{center}

\vspace{0.1in}

\Large{\bf Geometry from Brackets Consistency} \normalsize

\vspace{0.1in}

{\large \bf Edward Anderson$^*$}

\vspace{.2in}

\end{center}

\begin{abstract}

We argue for Brackets Consistency to be a `Pillar of Geometry', i.e.\ a foundational approach, other Pillars being 
1) Euclid's constructive approach, 
2) the algebraic approach, 
3) the projective approach, and 
4) the geometrical automorphism groups `Erlangen' approach.  
We proceed via a distinct Pillar 5) -- Killing's computational route to Erlangen's groups -- 
firstly by picking out the ensuing geometrically significant brackets subalgebras as consistent possibilities. 
Secondly, by considering which generators, arising piecemeal from different generalized Killing equations, combine brackets-consistently, 
recovering the affine--conformal and projective--conformal alternatives. 
Thirdly, we cease to rest on Killing's Pillar altogether, replacing it by mere polynomial ans\"{a}taze as source of generators. 
Even in this case -- demanding (Lie) Brackets Consistency -- 
suffices to recover notions of geometry, as indicated by the below alternative just dropping out of the ensuing brackets algebra, and is thus declared to be Pillar 6.
The current article is further motivated, firstly, by the gradually accumulating successes in reformulating Theoretical Physics 
to run on the Dirac Algorithm for classical constraint algebraic structure consistency.
Secondly, this is sufficiently driven by Brackets Consistency alone to transcend to whichever context possesses such brackets. 
Thirdly, we derive the conformal-or-projective alternative in top-automorphism-group $d \geq 3$ Flat Geometry.  
In a manner by now familiar in Dirac Algorithm applications, a) these two top geometries moreover arise side by side as roots of a merely algebraic quadratic equation. 
b) This equation arises from requiring that the self-bracket of a general ansatz for generators be strongly zero. 
c) For, if not, an infinite cascade of generators would follow, by which the ansatz would not support a finite Shape Theory.    
				
\end{abstract}

\n Mathematics keywords: Lie brackets. 
Algebraic closure. Foundations of Geometry.  Flat Geometry (including Conformal, Affine and Projective). Differential Geometry. Rigidity.  

\m 

\n Physics keywords: Constraints, constraint algebraic structures, Dirac Algorithm, Background Independence, Supersymmetry.  

\m 

\n PACS: 04.20.Cv, 02.40.-k 

\m 

\vspace{0.1in}
  
\n $^*$ Dr.E.Anderson.Maths.Physics *at* protonmail.com

\section{Introduction}\label{Intro-III}

\subsection{Four Pillars of Geometry}

\n In this Article, we reconsider the Foundations of Geometry, starting along the lines of John Stillwell's introductory account `The Four Pillars of Geometry' \cite{Stillwell}.  

\m

\n{\bf The First Pillar of Geometry} consists of constructions, starting with Euclid's `straightedge and compass' choice of instruments \cite{Elements}.  

\m 

\n{\bf The Second Pillar of Geometry} is algebraic approaches, from Cartesian Geometry \cite{Descartes, S04} 
                                                         to the vector, matrix, determinant and further methods \cite{Silvester} of Linear Algebra.

\m 														 
														 
\n{\bf The Third Pillar of Geometry} is Felix Klein's Erlangen program \cite{Klein, G67, Coxeter, S04}, 
based on transformation groups, leading to geometrically-significant automorphisms
\be 
Aut(\FrM, \bsigma)                                                             \m , 
\label{Aut}
\ee  
or the associated invariants under the action of these groups, 
\be 
{\cal I}nv(\sFrM, \bsigma)                                                  \m . 
\ee 
\n{\bf The Fourth Pillar of Geometry} is geometry from ray diagrams, of particular significance in formulating Projective Geometry \cite{Desargues, VY10, CG67, S04}, 
which Hilbert \cite{Hilb-Ax, HV32, Stillwell} showed in turn even have the capacity to axiomatize part of Algebra.

\end{titlepage}

\subsection{Given a list of distinct foundations for a subject...}

We moreover adopt a `Wheelerian' position as regards this basic list of foundational approaches.
This is named after John Archibald Wheeler's consideration of six routes to general Relativity \cite{Battelle, MTW}.
Namely that items on a list of foundational approaches are not only to be compared among themselves 
in structure assumed and extent of applicability and generalizability of each. 
Such lists are additionally to be added to, including as regards `Presupposition of Less Structure' 
                                               and finding `Zeroth Principles' for already-identified foundations' `First Principles' to rest upon.

\subsection{Fifth and Sixth Pillars of Geometry}

\n{\bf The Fifth Pillar of Geometry} is, very arguably along such lines \cite{PE-1, XIV, 9-Pillars}, 
the more advanced approach of Killing \cite{Killing} (as comprehensively generalized by Kentaro Yano \cite{Yano55, Yano70}). 
Here one solves a prescribed PDE -- the generalized Killing equation (GKE) 
\be 
{\cal GK} \, \u{\xi}  \:=  \pounds_{\u{\xi}} \bsigma = 0 
\ee 
for a given equipped manifold $\langle \FrM, \, \bsigma \rangle$. 
The solutions $\u{\xi}$ are generalized Killing vectors (GKVs), corresponding individually to automorphism generators 
                                                                          and collectively to the continuous automorphism group (\ref{Aut}), 
thus providing a {\sl systematic} way of computing this (see Sec 2 for examples). 
In this way, one arrives at the automorphism group end of Erlangen, 
so the `Killing' Fifth Pillar is a powerful \cite{Eisenhart33, Yano55, Yano70, MacCallum} Zeroth Principle for the `Erlangen' Third Pillar.   

\m 
 
\n{\bf The Sixth Pillar of Geometry} is argued in the current article to be {\it Brackets Consistency}. 
In this {\sl less structured} approach, one relies just on whether Lie brackets consistency 
\be
\mbox{\bf [}  c^{\st\sr\si\sa\sll}_A \mbox{\bf ,} c^{\st\sr\si\sa\sll}_B \mbox{\bf ]} = {C^C}_{AB} \, c^{\st\sr\si\sa\sll}_C  \m .  
\ee
holds for whatever candidate collection of generators one has at hand,   
The ${C^C}_{AB}$ here are structure constants.  
This has hitherto had a known but underused and underestimated status in the literature. 
The below motivational subsection outlines how the Theoretical Physics version of this approach -- Dirac's Algorithm \cite{Dirac} -- has gradually gained recognition.  
We perform our weakening in three steps, as follows. 

\m

\n Step 1) In Sec 3, we contemplate each specific flat space GKE's continuous automorphism group solution and {\sl pick out} its geometrically-significant consistent subgroups.
We do this in turn for the Euclidean, similarity, conformal \cite{O17}, affine \cite{G67} and projective \cite{S04} groups, 
noting the 1-$d$ simplifications and a 2-$d$ complication in the process.  

\m 

\n Step 2) In Sec 4, we next consider which generators $\u{\xi}^{\st\sr\si\sa\sll}$ {\sl obtained piecemeal from miscellaneous GKEs} are {\it mutually compatible}.  
We work within the shape-theoretically meaningful case of finitely-generated automorphism groups, 
noting the possible opposing phenomenon of infinite cascades of generators. 

\m 

\n Then on the one hand, we note that if we try to include a) special-conformal transformations 
alongside b) each of special-projective and affine transformations in $\geq 2$-$d$, a cascade is sourced.
In this way, a) and b) constitute an {\sl alternative}: only one of a) and b) can be concurrently accommodated in flat space.  

\m 

\n On the other hand, for $d = 1$ special-projective and special-conformal generators coincide up to sign, 
so conflation (rather than an alternative) ensues. 					 

\m

\n Step 3) In Sec 5, we finally cease to make any use of or reference to the machinery of Killing's approach \cite{Yano55, A-Killing}, 
replacing this by mere {\sl polynomial ans\"{a}taze} as sources of candidate generators $\u{c}^{\st\sr\si\sa\sll}$ after which we demand {\sl Brackets Consistency}: 
\be 
\mbox{\bf [}  c^{\st\sr\si\sa\sll}_A \mbox{\bf ,} c^{\st\sr\si\sa\sll}_B \mbox{\bf ]} \m \mbox{ finitely closes } \m .  
\ee   
\n We first consider the more straightforward 1-$d$ case, resulting in recovery of the above conflation.  

\m 

\n We next consider the $\geq 2$-$d$ case. 
This is now is found to encode the projective-or-conformal alternative as the two possible roots 
of a quadratic equation resultant from a bracket that has to strongly vanish in order to avoid sourcing a cascade.   

\m  

\n We finally deduce all of Sec 3's algebras within the more structurally sparse `Brackets Consistency' setting.
Secs 3 to 5 are supported by some simple structural considerations about splitting Lie algebras into blocks (Appendix A), 
including direct-product, semidirect-product and one-way integrability inter-relations between such blocks.

\subsection{Further motivation}  

The relational side of the Absolute versus Relational Debate \cite{Newton, L, M, DoD, Buckets, ABook} has led, firstly, to GR \cite{E16, Wald}, 
including via much later confirmation that GR's dynamical structure \cite{ADM, Dirac} is suitably relational \cite{BSW, B94I, RWR, AM13, ABook}.  
It has also lead to Background Independence \cite{A64, A67, Battelle, DeWitt67, Giu06, ABook, A-Killing, A-Cpct, A-CBI} parallels 
for most of the various currently proposed further Pillars of Geometry. 
The Problem of Time (as posed in \cite{DeWitt67, Battelle, K92, I93, APoT}) is also significant here, 
from failures to implement Background Independence constituting Problem of Time facets.

\m 

\n While $N$-point Shape Theory \cite{Kendall84, Kendall89, Small, Kendall, Bhatta, PE16, S-I, S-II, S-III, Minimal-N, A-Killing, A-Cpct} 
was initially developed toward a theory of Shape Statistics, 
this was also pointed out \cite{FORD, FileR, ABook} to serve as a simplified model of Background Independence 
This is moreover particularly close to the original Relational Mechanics setting \cite{BB82, FORD, FileR}, 
useful for small-concrete-$N$-Body Problems \cite{LR95, LR97, ML00, M02, M05, M15, Minimal-N}, 
and a model arena for classical and quantum Geometrodynamics \cite{K92, APoT2, ABook}.

\m 

\n Successfully formulating Background Independence -- and thus solving the Problem of Time, locally achieved in 
\cite{APoT2, APoT3, ABook, A-Letter, I, II, III, IV, V, VI, VII, VIII, IX, X, XI, XII, XIII, XIV} 
starts with Temporal and Configurational Relationalism principles formulated within the Relational side of the Absolute versus Relational Motion Debate.
It is however the third aspect of Background Independence -- Constraint Closure -- that the current Article latches onto.
As a Background Independence aspect, this is to be understood as the means by which the constraints {\sl produced} 
by each of Temporal and Configurational Relationalism are assessed for whether they are consistent. 
This moreover proceeds via a Dirac-type Algorithm \cite{Dirac, Sni, S82, HT92, AM13, ABook}, 
which is a somewhat more elaborate example of a Brackets Consistency approach, now for constraints $\u{\scC}$ plugged into Poisson brackets, 
\be 
\mbox{\bf \{} \, \scC_A \mbox{\bf ,} \, \scC_B \, \mbox{\bf \}} \es {C^C}_{AB} \, \scC_C                    \m .   
\label{CC}
\ee
This approach moreover leads to number of further successes. 
These include recovery of both \cite{RWR, AM13, ABook} GR dynamical features and even universal locally Lorentzian relativity of SR 
from Presupposing Less Structure, by which it solves the classical version of a further Problem of Time facet: 
the Spacetime Reconstruction Problem \cite{Battelle, K92, I93}.
This result is based on a key constraints bracket providing a polynomial prefactor times a term which either produces an infinite cascade 
or violates a further Background Independence aspect: Refoliation Invariance. 
So one way to attain consistency is for the polynomial prefactor to be zero. 
And this moreover embodies the Galilean versus Lorentzian fork for the local form the universal Relativity, 
now as a mere matter of which root of this polynomial one selects in order to attain the strong vanishing.

\m 

\n The current Article's approach to Geometry is then more concretely motivated as follows. 

\m 

\n A) Successes in in reformulating classical Theoretical Physics to run on the Dirac Algorithm -- for classical constraint algebraic structure consistency -- 
have been piling up. 

\m 

\n B) These successes are sufficiently driven by `mere brackets consistency' preserving rigidity \cite{IX} transcending to \cite{XIV} whichever context has such brackets. 

\m 

\n C) In a manner by now familiar in applications of the Dirac Algorithm \cite{RWR, OM02, OM03, Phan, AM13, ABook}, 
we derive the conformal-or-projective ambiguity in top-group Flat Geometry  
as the two possible strongly vanishing roots of a brackets relation for the bracket both of whose entries take the general ansatz for a second-order generator.  
Brackets Consistency thus has the capacity to recover notions of geometry, 
as indicated by the above ambiguity now just dropping out of the ensuing brackets algebra.

\m 

\n The Conclusion (Sec 6) moreover mentions further parallels between Background Independence aspects and Foundations of Geometry Pillars.

\section{Killing's approach: generalized Killing equations producing geometrical automorphism groups as complete packages}

\n Let $\bFrM$ be a differentiable manifold.  
Consider thereupon an infinitesimal transformation 
\be 
\u{x} \longrightarrow \u{x}^{\prime} + \epsilon \, \u{\xi} \m .
\label{inf}
\ee 
For this to preserve an object, say a tensor, $\bT$, substituting (\ref{inf}) into $\bT$'s transformation law and equating first-order terms in $\epsilon$ gives 
\be 
\pounds_{\u{\xi}}\bT = 0 \m .  
\ee
for $\pounds_{\u{\xi}}$ the Lie derivative \cite{Yano55} with respect to $\u{\xi}$: a differential operator guaranteed by $\bFrM$'s differential structure. 
Below we consider a fortiori geometrically significant objects in the role of $\bT$. 

\m 

\n{\bf Definition 1} For $\langle \, \bFrM, \, \bsigma \, \rangle$ a manifold $\bFrM$ equipped with a geometrically-significant level of structure $\bsigma$, 
the {\it generalized Killing equation (GKE)} is 
\be 
{\cal GK} \, \u{\xi} :=  \pounds_{\u{\xi}} \bsigma  =  0  \m . 
\ee 
This is moreover to be regarded as a PDE to solve for $\u{\xi}$, such solutions being termed {\it generalized Killing vectors (GKV)}.  
These form the algebra corresponding to the continuous part of the automorphism group $Aut(\bFrM, \bsigma)$.    

\m 

\n{\bf Example 1} The most familiar case is as follows. 
For  $\langle \, \bFrM, \, \bg \, \rangle$ an (arbitrary-signature) Riemannian manifold, i.e.\ carrying a Riemannian metric structure $\bg$, 
\be 
{\cal K} \, \u{\xi} :=  \pounds_{\u{\xi}} \bg = 0  \m 
\label{LKE}
\ee   
is {\it Killing's equation}. 
Its solutions $\u{\xi}$ are {\it Killing vectors}.  
These are of significance in the study of  $\langle \, \bFrM, \, \bg \, \rangle$ as they correspond to the {\sl symmetries}, 
more precisely to the {\sl isometries}, the totality of Killing vectors forming the {\it isometry group} 
\be 
Isom(\bFrM, \, \bg )  \m .
\ee
\n{\bf Example 2} For  $\left\langle \, \bFrM, \, \overline{\bg} \, \right\rangle$ a manifold equipped with metric structure modulo constant rescalings, $\overline{\bg}$,
\be 
{\cal SK} \u{\xi}  :=  \pounds_{\u{\xi}} \overline{\bg} = 0 
\label{LSKE}
\ee 
is the {\it similarity Killing equation} \cite{MacCallum} (alias {\it homothetic Killing equation} in \cite{Yano55}).  
Its solutions $\u{\xi}$ are {\it similarity Killing vectors}, 
consisting of isometries alongside 1 or 0 {\it proper similarities}, i.e.\ non-isometries.   
The totality of these form the {\it similarity group} 
\be 
Sim \left( \bFrM, \, \overline{\bg} \right)                                             \m ;  
\ee 
for $\langle \, \bFrM, \, \bg \, \rangle$ possessing 0 proper similarities, 
\be
Sim \left( \bFrM, \, \overline{\bg} \right) = Isom \left( \bFrM, \, \bg \right)  \m .  
\ee
\n{\bf Example 3} For  $\left\langle \, \bFrM, \, \widetilde{\bg} \, \right\rangle$ a manifold equipped with metric structure modulo local rescalings, $\w{\bg}$, 
\be 
{\cal CK} \, \u{\xi} :=  \pounds_{\u{\xi}} \widetilde{\bg} = 0 
\label{LCKE}
\ee 
is the {\it conformal Killing equation}.  
Its solutions $\u{\xi}$ are {\it conformal Killing vectors}, 
consisting of similarities and special conformal transformations, 
except in `$\mathbb{R}^2$ or $\mathbb{C}$' for which we get an infinity of analytic functions 
due to the  flat CKE collapsing in 2-$d$ to the Cauchy--Riemann equations \cite{AMP}.
The totality of CKVs form the {\it conformal group}
\be 
Conf( \bFrM, \, \w{\bg} )  \m .  
\ee  
\n{\bf Example 4} For  $\langle \, \bFrM, \, {\bGamma} \, \rangle$ a manifold equipped with affine structure,  $\bGamma$
\be 
{\cal AK} \, \u{\xi} :=  \pounds_{\u{\xi}} \bGamma = 0 
\label{LAKE}
\ee 
is the {\it affine Killing equation}. 
Its solutions $\u{\xi}$ are {\it affine Killing vectors}, consisting of similarities alongside shears and squeezes alias Procrustes stretches \cite{Coxeter}.  
The totality of these form the {\it affine group} 
\be 
Aff( \bFrM, \, \bGamma )            \m . 
\ee 
\n{\bf Example 5} For  $\left\langle \, \bFrM, \, \s{P}{\bGamma} \, \right\rangle$ a manifold equipped with projective structure,  

\n\be 
{\cal PK} \, \u{\xi}  := \pounds_{\u{\xi}} \s{P}{\bGamma} = 0 
\label{LPKE}
\ee 
is the {\it projective Killing equation}.    
Its solutions $\u{\xi}$ are {\it projective Killing vectors}, consisting of affine transformations alongside 
`special projective transformations'.    
The totality of these form the {\it projective group} 

\n\be 
Proj\left( \bFrM, \, \s{P}{\bGamma} \right)           \m . 
\ee 
Having built up these example, we entertain some discussion on nomenclature. 
Killing found the first such equation; its solutions form the isometry group of a geometry. 
{\it Generalized Killing equation (GKE)} is then a collective name for the five equations (\ref{LKE}, \ref{LSKE}, \ref{LCKE}, \ref{LAKE}, \ref{LPKE}) (and more \cite{Yano55}), 
and  {\it generalized Killing vectors (GKV)} for the corresponding solutions.   
A more conceptually descriptive name is, moreover, {\it automorphism equation}, as explained in \cite{PE-1, XIV, 9-Pillars}, 
whose solutions are now of course just termed {\it automorphism generators}.

\section{Geometrically-significant consistent subgroups within the Killing approach}

\n Let us start `from the top' by solving some particular GKEs for entire automorphism algebras. 
We then pick out subalgebras by inspecting its brackets relations for which combinations of generators automatically close.

\subsection{$Eucl(d)$}

\n{\bf Remark 1} Solving the flat Killing equation \cite{Yano55, Yano70, AMP, MacCallum} gives the isometry group generators  
\be 
P_A := \pa_A \m \mbox{ (translations) }
\label{P-def}
\ee 
and 
\be 
\Omega_{AB} := x_A\pa_B - x_B\pa_A  \m \mbox{ (rotations) }                                                                                                    \m . 
\label{O-def}
\ee
\n{\bf Structure 1} We thus have the 2-block (in the sense of Appendix A) Euclidean algebra $Eucl(d)$.
\be 
\mbox{\bf [}  P_A \mbox{\bf ,} \, P_{B}  \mbox{\bf ]}                =  0                                                                        \m , 
\label{P-P}
\ee 
\be 	 
\mbox{\bf [}  P_A \mbox{\bf ,} \, \Omega_{BC}  \mbox{\bf ]}          =  2 \, \delta_{A[B} P_{C]}                                                 \m , 
\label{P-O}
\ee 
\be 	 
\mbox{\bf [}  \Omega_{AB} \mbox{\bf ,} \, \Omega_{CD}  \mbox{\bf ]}  \es  2 \left( \delta_{C[B}\Omega_{A]D} + \delta_{D[B} \Omega_{A]C} \right)  \m .  
\label{O-O}
\ee 
\n{\bf Remark 2} The first bracket means that the $\u{P}$ mutually commute.

\m

\n The third bracket demonstrates that the rotations form the standard special-orthogonal algebra, 
\be 
Rot(d) = SO(d) \m . 
\label{Rot-S}
\ee 
\n Finally, the second bracket signifies that $\u{P}$ is a $SO(d)$-vector. 

\m 

\n{\bf Remark 3} These brackets have the overall semidirect product structure (c.f.\ Appendix A)
\be 
Eucl(d) \es         Tr(d) \rtimes Rot(d)  
        \es  \mathbb{R}^d \rtimes SO(d)                                                                                                          \m . 
\ee 
We can also see that 
\be 
Tr(d) = \mathbb{R}^d
\ee 
and (\ref{Rot-S}) each close; these are geometrically significant subalgebras, and form the bounded lattice \cite{Lattice} of Fig \ref{Bra-Eucl}.b).   
%
{            \begin{figure}[!ht]
\centering
\includegraphics[width=0.4\textwidth]{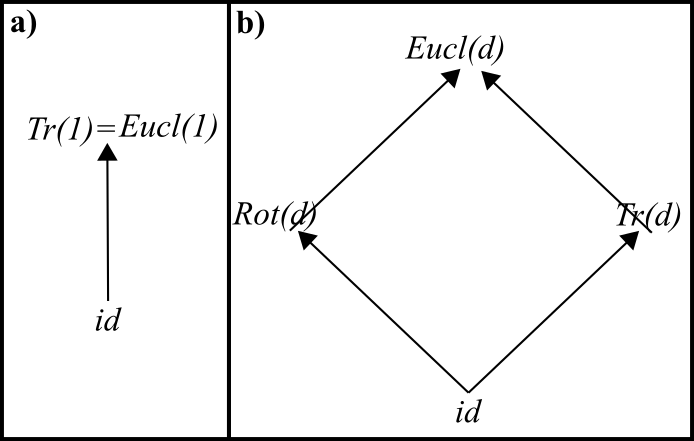}
\caption[Text der im Bilderverzeichnis auftaucht]{        \footnotesize{Geometrically-significant subgroups of $Eucl(d)$ a) in 1-$d$ and b) in $\geq 2$-$d$.}}
\label{Bra-Eucl} \end{figure}          }

\m 

\n{\bf Remark 4} In 1-$d$, the Euclidean algebra collapses by the lack of continuous rotations to just 
\be 
\mbox{\bf [}  P \mbox{\bf ,} \, P  \mbox{\bf ]}                =  0                                                                    \m ;
\label{P-P-1-d}
\ee 
i.e.\  
\be 
Eucl(1) = Tr(1) = \mathbb{R}                                                                                                                   \m . 
\ee 
See Fig \ref{Bra-Eucl}.a) for the more chain of geometrically significant subgroups in this case.

\subsection{$Sim(d)$}

\n{\bf Remark 1} Solving the flat similarity Killing equation \cite{Yano55, Yano70, MacCallum} 
gives not only the flat Killing equation solution but also an extra generator:    
\be 
D = x^A \pa_A     \m \mbox{ (dilations) }                                                                                                             \m .  
\label{D-Def}
\ee 
\n{\bf Structure 1} We thus have the following 3-block similarity algebra $Sim(d)$.  
\be 
\mbox{\bf [}  P_A \mbox{\bf ,} \, P_{B}  \mbox{\bf ]}                =  0                                                                             \m , 
\label{P-P-S}
\ee 
\be 	 
\mbox{\bf [}  P_A \mbox{\bf ,} \, \Omega_{BC}  \mbox{\bf ]}          =  2 \, \delta_{A[B} P_{C]}                                                      \m , 
\label{P-O-S}
\ee
\be 	 
\mbox{\bf [}  \Omega_{AB} \mbox{\bf ,} \, \Omega_{CD}  \mbox{\bf ]}  \es  2 \left(  \delta_{C[B}\Omega_{A]D} + \delta_{D[B} \Omega_{A]C} \right)      \m , 
\label{O-O-S}
\ee 
\be 	 
\mbox{\bf [}  D   \mbox{\bf ,} \, D  \mbox{\bf ]}                    =  0                                                                             \m , 
\label{D-D}
\ee
\be 	 
\mbox{\bf [}  P_A \mbox{\bf ,} \,  D  \mbox{\bf ]}                   =  P_A                                                                           \m , 
\label{P-D}
\ee 
\be 	 
\mbox{\bf [}  D   \mbox{\bf ,} \, \Omega_{AB}  \mbox{\bf ]}          =  0                                                                             \m .   
\label{D-O}
\ee
\n{\bf Remark 2} The first three brackets inherit their meanings from the preceding section.  

\m 

\n The fourth bracket just means that single-generator blocks necessarily self-commute. 

\m 

\n The fifth bracket signifies that $\u{P}$ is a $Dil$-vector. 

\m 

\n Finally, the sixth bracket encodes a direct product whose significance is dilation--rotation independence.     

\m 

\n\n{\bf Remark 3} The similarity brackets algebra has the overall structure of a semidirect product of a direct product, more specifically 
\be 
Sim(d) \es  Tr(d) \rtimes (Rot(d) \times Dil)  
       \es  \mathbb{R}^d \rtimes (SO(d) \times \mathbb{R}_+)                                                                                          \m .  
\ee 
\n{\bf Remark 4} We can also read off that all single blocks -- $Tr(d)$, $Dil$ and $Rot(d)$ -- 
                            and pairs -- $Eucl(d)$, 
\be 
\mbox{ (dilatations) } \m Dilatat(d)  \es  Tr(d) \rtimes Dil  
                                      \es  \mathbb{R}^d \rtimes \mathbb{R}_+                                                                          \m ,
\ee  
and 
\be 
Rot(d) \times Dil                     \es  SO(d) \times \mathbb{R}_+ 
\ee 
-- stand alone as subalgebras, as well as $Sim(d)$'s triple combination being consistent.  

\m 

\n These form the bounded lattice of subalgebras of Fig \ref{Bra-Sim}.b).    
%
{            \begin{figure}[!ht]
\centering
\includegraphics[width=0.6\textwidth]{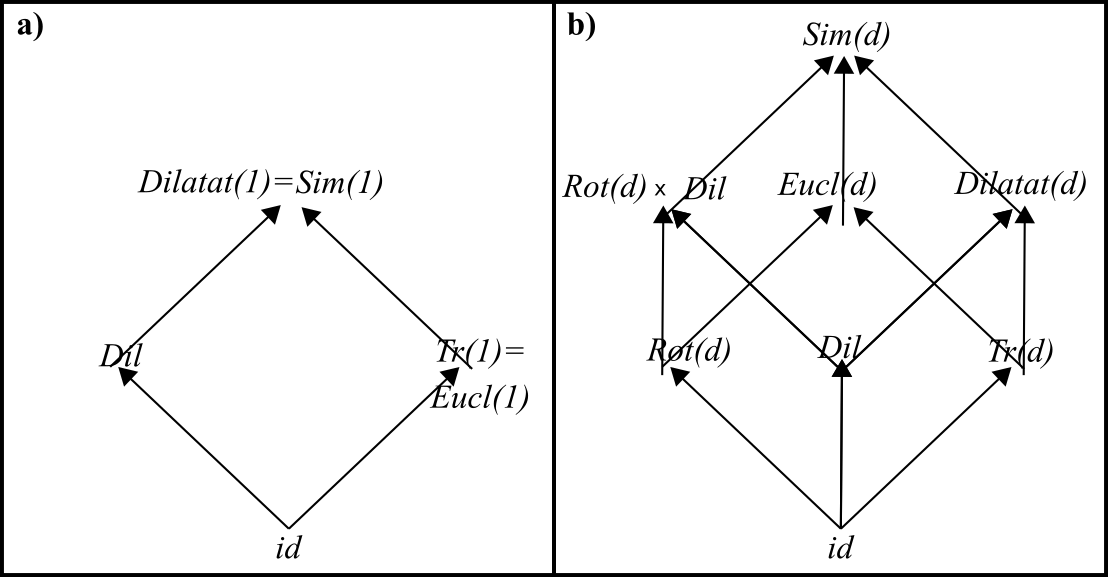}
\caption[Text der im Bilderverzeichnis auftaucht]{        \footnotesize{Geometrically-significant subgroups of $Sim(d)$ a) in 1-$d$ and b) in $\geq 2$-$d$.}}
\label{Bra-Sim} \end{figure}          }

\m 

\n\n{\bf Remark 5} In 1-$d$, the $Sim$ algebra collapses to 
\be 
\mbox{\bf [} P \mbox{\bf ,} \, P \mbox{\bf ]} = 0      \m ,
\label{P-P-1-d-Dilatat}
\ee 
\be 
\mbox{\bf [} P \mbox{\bf ,} \, D \mbox{\bf ]} = P      \m ,
\label{P-D-1-d}
\ee
\be   
\mbox{\bf [} D \mbox{\bf ,} \, D \mbox{\bf ]} = 0      \m , 
\label{D-D-1-d}
\ee
forming the 1-$d$ dilatations.
This is in the form of a semidirect product of 1-$d$ blocks (c.f.\ Appendix A), more specifically  
\be 
Sim(1) \es  Dilatat(1)  
       \es  \mathbb{R} \rtimes \mathbb{R}_+            \m . 
\ee 
Its smaller repertoire of geometrically-significant subalgebras can be found in Fig \ref{Bra-Sim}.a).

\subsection{The $Conf(d)$ family}

{\bf Remark 1} Solving the flat conformal Killing equation \cite{Yano55, Yano70, AMP} 
gives not only the flat similarity Killing equation but an extra generator,   
\be 
K^A  :=  ||x||^2\pa^A - 2 \, x^A x^B \pa_B   \m ,
\label{K-Def}
\ee 
or, in operator form, 
\be 
K^A \es \left(  \delta^{AB}||x||^2 - 2 \, x^A x^B  \right) \pa_B                                                                                       \m .  
\label{K-Def-Op}
\ee 
These are known as {\it special conformal transformation} generators \cite{O17}.

\m 

\n By the comment in Sec 2, for $d \geq 3$ this is a complete description, whereas $d = 2$ has a further infinite set of generators.  

\m

\n{\bf Structure 1} For $d \geq 3$, we thus have the following 4-block conformal algebra $Conf(d)$.  
\be 
\mbox{\bf [}  P_A \mbox{\bf ,} \, P_{B}  \mbox{\bf ]}                =  0                                                                             \m , 
\label{P-P-C}
\ee 
\be 	 
\mbox{\bf [}  P_A \mbox{\bf ,} \,  D  \mbox{\bf ]}                   =  P_A                                                                           \m , 
\label{P-D-C}
\ee 
\be 	 
\mbox{\bf [}  P_A \mbox{\bf ,} \, \Omega_{BC}  \mbox{\bf ]}          =  2 \, \delta_{A[B} P_{C]}                                                      \m , 
\label{P-O-C}
\ee 
\be 	 
\mbox{\bf [}  D   \mbox{\bf ,} \, D  \mbox{\bf ]}                    =  0                                                                             \m , 
\label{D-D-C}
\ee 
\be 	 
\mbox{\bf [}  D   \mbox{\bf ,} \, \Omega_{AB}  \mbox{\bf ]}          =  0                                                                             \m , 
\label{D-O-C}
\ee 
\be 	 
\mbox{\bf [}  \Omega_{AB} \mbox{\bf ,} \, \Omega_{CD}  \mbox{\bf ]}  \es  2 \left( \delta_{C[B}\Omega_{A]D} + \delta_{D[B} \Omega_{A]C} \right)       \m , 
\label{O-O-C}
\ee
\be 
\mbox{\bf [}  K_A \mbox{\bf ,} \, K_{B}  \mbox{\bf ]}                =  0                                                                             \m ,   
\label{K-K}
\ee
\be 
\mbox{\bf [}  K_A \mbox{\bf ,} \, \Omega_{BC}  \mbox{\bf ]}          =  2 \, \delta_{A[B} P_{K]}                                                      \m ,
\label{K-O}
\ee
\be 
\mbox{\bf [}  K_A \mbox{\bf ,} \, D  \mbox{\bf ]}                    =  - K_A                                                                         \m ,
\label{K-D}
\ee
\be 
\mbox{\bf [}  K_A \mbox{\bf ,} \, P_B  \mbox{\bf ]}                 \es  2 \left( \Omega_{AB} + \delta_{AB} D \right)                                \m .   
\label{K-P}
\ee
{\bf Remark 2} Six of these brackets are to be interpreted as in $Sim(d)$. 

\m 

\n The seventh bracket means that the special conformal transformations $\u{K}$ mutually commute. 

\m 

\n The eighth bracket signifies that the $\u{K}$ are $SO(d)$-vectors.

\m 

\n The ninth bracket means that the $\u{K}$ are dilational covectors. 

\m 

\n Finally, the tenth bracket encodes the one-way mutual integrability 
\be 
(\u{P}, \u{K}) \, \, \Thomas \, \, (D, \u{\u{\Omega}})                                                                                                            \m .  
\label{PK->DO}
\ee 
After a bit of algebra, these brackets demonstrate that the totality of $Conf(d)$ for $d \geq 3$ constitutes 
\be 
Conf(d)  \es  SO(d + 1, 1) \m : \m \mbox{ the $d + 1$ dimensional Lorentz group} \m .  
\label{Conf(d)}
\ee 
{\bf Remark 3} See Fig \ref{Bra-Conf} for the bounded lattice of geometrically meaningful subalgebras of $Conf(d)$ $d \geq 3$; 
the above integrability is crucial in limiting which combinations of generators close consistently.
%
{            \begin{figure}[!ht]
\centering
\includegraphics[width=0.4\textwidth]{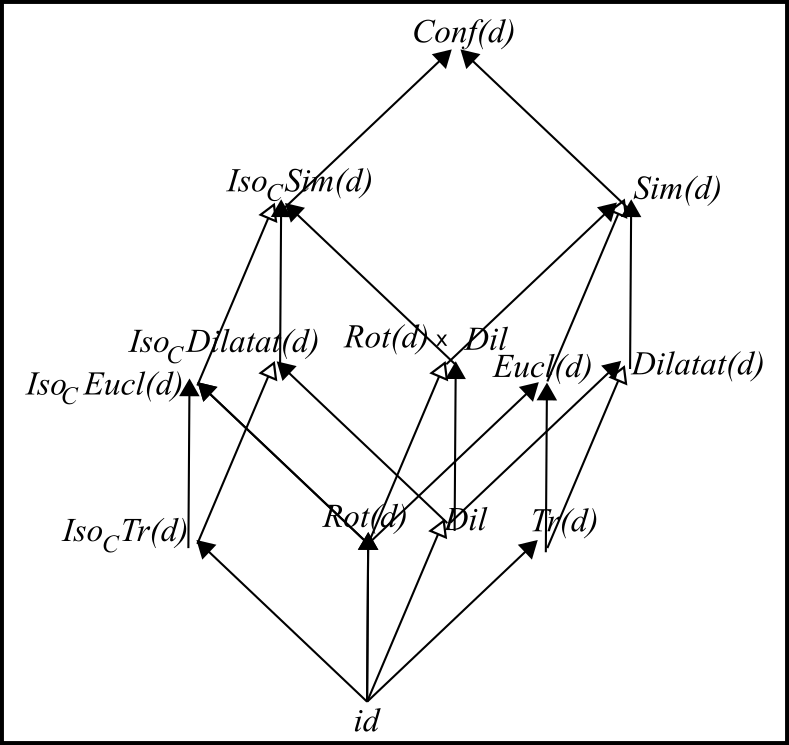}
\caption[Text der im Bilderverzeichnis auftaucht]{        \footnotesize{Geometrically-significant subgroups of $Conf(d)$ a) in 1-$d$ and b) in $\geq 2$-$d$.}}
\label{Bra-Conf} \end{figure}          }

\m 

\n{\bf Remark 4} In particular, $C\mbox{-}Iso\mbox{-}Sim(d)$ consists of $\u{K}$, $D$ and $\u{\u{\Omega}}$, closing according to  
\be 
\mbox{\bf [}  K_A \mbox{\bf ,} \, K_{B}  \mbox{\bf ]}                =  0                                                                             \m ,   
\ee
\be 
\mbox{\bf [}  K_A \mbox{\bf ,} \, \Omega_{BC}  \mbox{\bf ]}          =  2 \, \delta_{A[B} P_{K]}                                                      \m ,
\ee
\be 
\mbox{\bf [}  K_A \mbox{\bf ,} \, D  \mbox{\bf ]}                    =  - K_A                                                                         \m ,
\ee
\be 	 
\mbox{\bf [}  D   \mbox{\bf ,} \, D  \mbox{\bf ]}                    =  0                                                                             \m , 
\ee 
\be 	 
\mbox{\bf [}  D   \mbox{\bf ,} \, \Omega_{AB}  \mbox{\bf ]}          =  0                                                                             \m , 
\ee 
\be 	 
\mbox{\bf [}  \Omega_{AB} \mbox{\bf ,} \, \Omega_{CD}  \mbox{\bf ]}  \es  2 \left( \delta_{C[B}\Omega_{A]D} + \delta_{D[B} \Omega_{A]C} \right)       \m .   
\ee
That this is indeed isomorphic to $Sim(d)$, can be ascertained by setting 
\be 
-D = E \m \mbox{ in the } \m Iso \m \mbox{ case and } \m D = E \m \mbox{ in the usual algebra itself } \m ; 
\label{E-Move}
\ee 
the brackets are then in 1 : 1 correspondence. 
This isomorphism is moreover inherited under taking subgroups.

\subsection{$Aff(d)$}

\n{\bf Remark 1} Solving the flat affine Killing equation \cite{Yano55, Yano70}, we arrive at the unsplit $GL(d, \mathbb{R})$ {\it generators}  
\be 
{G^A}_B   :=  x^A \pa_B \m , 
\label{G-Def}
\ee
\n{\bf Structure 1} We thus have the following 2-block rendition of the affine algebra $Aff(d)$. 
\be 	 
\mbox{\bf [}  P_A     \mbox{\bf ,} \, P_B      \mbox{\bf ]}  =  0                                    \m ,  
\label{P-P-AU}
\ee 
\be 	 
\mbox{\bf [}  P_C     \mbox{\bf ,} \, {G^A}_B  \mbox{\bf ]}  =  {\delta^A}_C P_B                     \m , 
\label{P-G}
\ee 
\be 	 
\mbox{\bf [}  {G^A}_B \mbox{\bf ,} \, {G^C}_D  \mbox{\bf ]}  =  2 \, {\delta^{[C}}_B {G^{A]}}_D      \m . 
\label{G-G}
\ee 	
{\bf Remark 2} The third bracket is the $GL(d, \mathbb{R})$ algebra itself.
 
\m 
 
\n The second bracket means that $\u{P}$ is a $GL(d, \mathbb{R})$ vector.  

\m 

\n{\bf Remark 3} We have overall the semidirect product structure 
\be 
Aff(d)  \es  \mathbb{R}^d \rtimes GL(d, \mathbb{R})                                                   \m ,  
\ee 
with the $Tr(d) = \mathbb{R}^d$ algebra of the first bracket and the $GL(d, \mathbb{R})$ 
algebra of the third bracket as immediately identifiable and geometrically significant subalgebras.  

\m 

\n{\bf Remark 4} $d \geq 2$ is necessary in order to support a distinct affine algebra.  
%
{            \begin{figure}[!ht]
\centering
\includegraphics[width=0.7\textwidth]{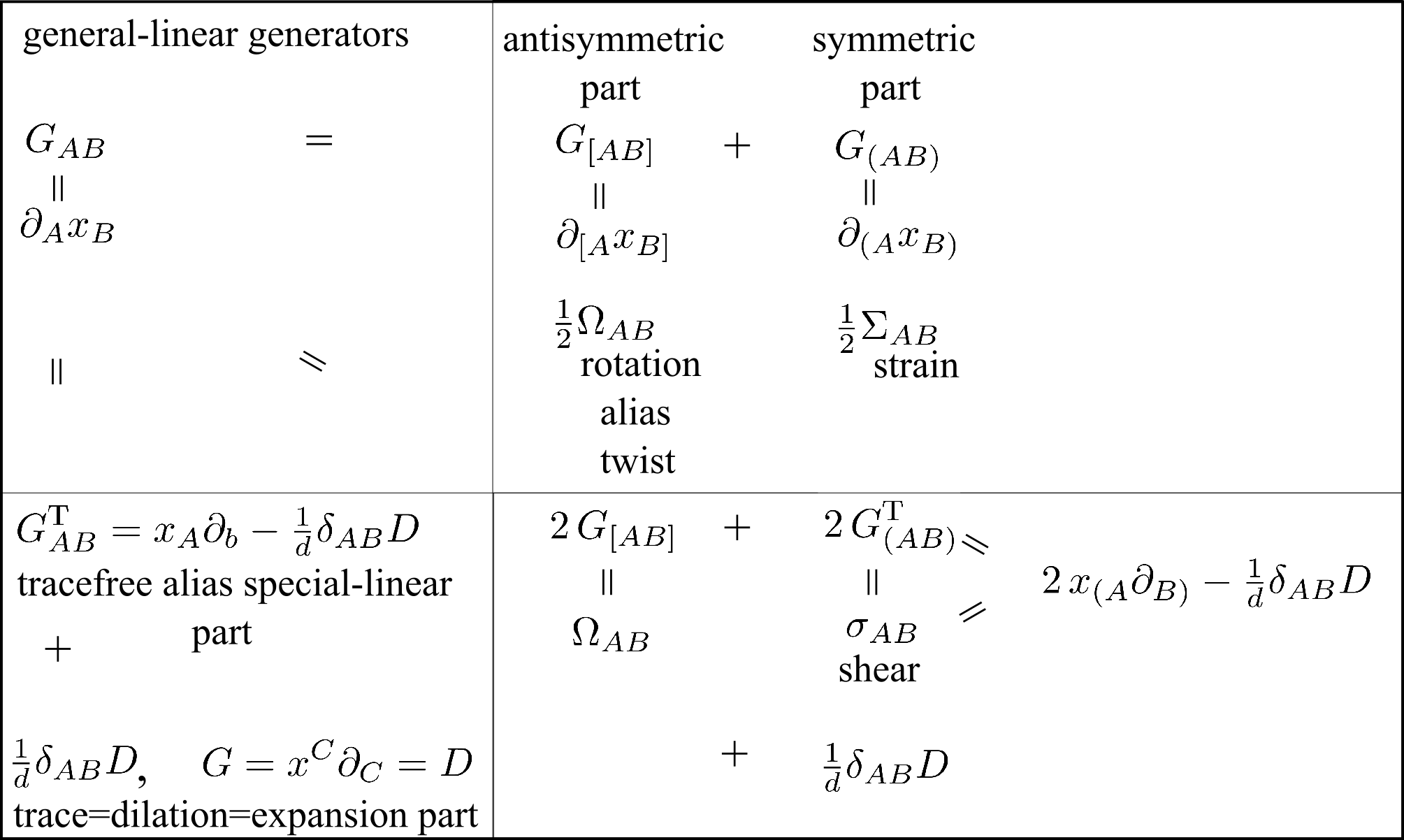}
\caption[Text der im Bilderverzeichnis auftaucht]{        \footnotesize{The symmetric-antisymmetric and trace-tracefree splits, 
including the application of both at once, with irreducible representations identified.}}
\label{Splits} \end{figure}          }

\m 

\n{\bf Remark 5} Finding further geometrically significant algebras benefits from making 
one or both of the symmetric--antisymmetric and trace--tracefree splits of ${G^A}_B$ of Fig \ref{Splits}.  

\m 

\n Introducing the {\it shear generator}
\be 
\sigma_{AB}  \:= 2  \, x_{(A}\pa_{B)}  \m - \m  \frac{1}{d} \, \delta_{AB} D                                                                            \m ,  
\label{S-Def}
\ee 
we have the following 4-block rendition of the affine algebra $Aff(d)$. 
\be 
\mbox{\bf [}  P_A \mbox{\bf ,} \, P_{B}  \mbox{\bf ]}                =  0                                                                            \m , 
\label{P-P-A}
\ee
\be 	 
\mbox{\bf [}  P_A \mbox{\bf ,} \, \Omega_{BC}  \mbox{\bf ]}          =  2 \, \delta_{A[B} P_{C]}                                                     \m , 
\label{P-O-A}
\ee
\be 	 
\mbox{\bf [}  \Omega_{AB} \mbox{\bf ,} \, \Omega_{CD}  \mbox{\bf ]}  \es  2 \left( \delta_{C[B}\Omega_{A]D} \m + \m  \delta_{D[B} \Omega_{A]C} \right)      \m , 
\label{O-O-A}
\ee
\be 	 
\mbox{\bf [}  D   \mbox{\bf ,} \, D  \mbox{\bf ]}                    =  0                                                                            \m , 
\label{D-D-A}
\ee 
\be 	 
\mbox{\bf [}  P_A \mbox{\bf ,} \,  D  \mbox{\bf ]}                   =  P_A                                                                          \m , 
\label{P-D-A}
\ee 
\be 	 
\mbox{\bf [}  D   \mbox{\bf ,} \, \Omega_{AB}  \mbox{\bf ]}          =  0                                                                            \m , 
\label{D-O-A}
\ee 
\be 	 
\mbox{\bf [}  P_A \mbox{\bf ,} \, \sigma_{BC}  \mbox{\bf ]}         \es  2 \, \delta_{A(B} P_{C)}  \m - \m \frac{1}{d} \, \delta_{BC} P_A               \m , 
\label{P-S}
\ee 
\be 	 
\mbox{\bf [}  D , \mbox{\bf \,} \sigma_{AB}  \mbox{\bf ]}            =  0                                                                            \m , 
\label{D-S}
\ee 
\be 	 
\mbox{\bf [}  \sigma_{AB} \mbox{\bf ,} \, \Omega_{CD}  \mbox{\bf ]}  \es  2 \left( \delta_{C(B} \sigma_{A)D} - \delta_{D(B} \sigma_{A)C} \right)     \m , 
\label{S-O}
\ee
\be 	 
\mbox{\bf [}  \sigma_{AB} \mbox{\bf ,}  \, \sigma_{CD} \mbox{\bf ]}  =  2 \left( \delta_{D(B} \Omega_{A)C} + \delta_{C(B} \Omega_{A)D} \right)       \m .   
\label{S-S}
\ee
{\bf Remark 6} The first six of these brackets are to be interpreted as in Sec 6, for $Aff(d)$ is another extension of $Sim(d)$. 

\m 

\n The seventh bracket signifies that $\u{P}$ extends from being a $SO(d)$-vector to being a $SL(d, \mathbb{R})$-vector

\m 

\n The eighth bracket means dilation--shear independence.

\m 

\n The ninth bracket signifies that $\u{\u{\sigma}}$ is a $SO(d)$-tensor. 

\m 

\n Finally, the tenth bracket encodes the 1-way self integrability 
\be 
\u{\u{\sigma}} \, \, \Thomas \, \,  \u{\u{\Omega}}                                                                                                                \m .
\label{S->O}
\ee 
\n{\bf Remark 7} The overall structure of the affine algebra is again the semidirect product of a direct product.  
More specifically
\be 
Aff(d)  \es  Tr(d)         \rtimes (SL(d, \mathbb{R}) \times Dil)  
        \es  \mathbb{R}^d  \rtimes (SL(d, \mathbb{R}) \times \mathbb{R}_+)  \m , 
\ee 
the 1-way self-integrability (\ref{S->O}) remaining hidden within the $SL(d, \mathbb{R})$ 
double block of the $\u{\u{\omega}}$ and $\u{\u{\sigma}}$. 

\m 

\n{\bf Remark 8} The lattice of geometrically-significant subalgebras of Fig \ref{Bra-Aff} can be read off our brackets algebra.  
%
{            \begin{figure}[!ht]
\centering
\includegraphics[width=0.4\textwidth]{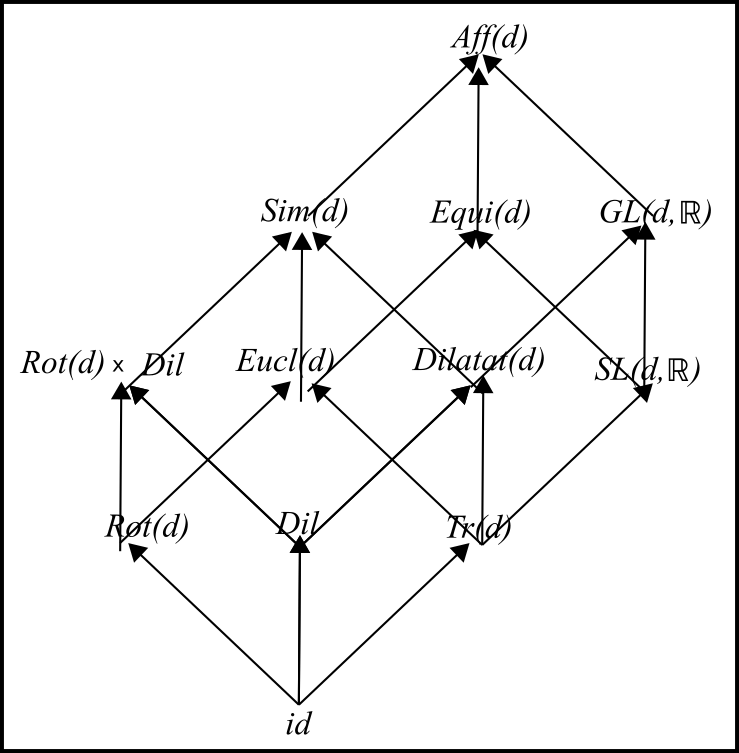}
\caption[Text der im Bilderverzeichnis auftaucht]{        \footnotesize{Geometrically-significant subgroups of $Aff(d)$ a) in 1-$d$ and b) in $\geq 2$-$d$. 
$Equi(d)$ is the equi-top-form group, corresponding to equi-top-form geometry, e.g.\ equiareal geometry in 2-$d$ \cite{Coxeter}.  }}
\label{Bra-Aff} \end{figure}          }

\subsection{The $Proj(d)$ family}

\n{\bf Remark 1} Solving the projective Killing equation \cite{Yano55, Yano70} returns the affine generators and, additionally, 
the {\it special projective transformation} generators,  
\be 
Q^A   :=  x^A x^B \pa_B \m . 
\ee  
\n{\bf Structure 1} Together, these form the following 3-block rendition of the projective algebra $Proj(d)$. 
\be 	 
\mbox{\bf [}  P_A     \mbox{\bf ,} \, P_B      \mbox{\bf ]}  =  0                                         \m ,  
\label{P-P-PU}
\ee 
\be 	 
\mbox{\bf [}  P_C     \mbox{\bf ,} \, {G^A}_B  \mbox{\bf ]}  =   {\delta^A}_C P_B                         \m , 
\label{P-G-PU}
\ee 
\be 	 
\mbox{\bf [}  {G^A}_B \mbox{\bf ,} \, {G^C}_D  \mbox{\bf ]}  =  2 \, {\delta^{[C}}_B {G^{A]}}_D           \m , 
\label{G-G-PU}
\ee 	 
\be 	 
\mbox{\bf [}  Q^A     \mbox{\bf ,} \, Q^B      \mbox{\bf ]}  =  0                                         \m ,  
\label{Q-Q-2}
\ee 
\be 	 
\mbox{\bf [}  Q^C     \mbox{\bf ,} \, {G^A}_B  \mbox{\bf ]}  =  - {\delta^C}_B Q^A                        \m ,  
\label{Q-G}
\ee
\be 	 
\mbox{\bf [}  P_A     \mbox{\bf ,} \, Q^B  \mbox{\bf ]}  =  2 \, {\delta_A}^{(B} {\delta_C}^{D)} {G^C}_D  \m .
\label{P-Q}
\ee 
\n{\bf Remark 2} The first three of these brackets have the same interpretation as in the preceding section. 

\m 

\n The fourth bracket signifies that the special projective transformations $\u{Q}$ mutually commute. 

\m 

\n The fifth bracket means that $\u{Q}$ is a $GL(d, \mathbb{R})$-vector.

\m 

\n Finally, the sixth bracket encodes the 1-way mutual integrability 
\be 
(\u{P}, \u{Q}) \, \, \Thomas \, \, \u{\u{G}}   \m . 
\label{PQ->G}
\ee 
\n{\bf Remark 3} Together, these form 
\be 
Proj(d)  \es  PGL(d + 1, \mathbb{R}) \m :
\ee
the $d$-dimensional real projective general linear algebra. 

\m 

\n{\bf Remark 3} Aside from a consistent $Aff(d)$ subalgebra and subalgebras thereof, 
this brackets algebraic structure contains a 'second version' of everything with $\u{Q}$ in place of $\u{P}$. 
I.e.\ we start to have lesserly known geometrically significant subalgebras in this case. 
Moreover if both $\u{Q}$ and $\u{P}$ are present, $\u{\u{G}}$ cannot be dropped by the 1-way integrability (\ref{PQ->G}).  
The translational algebra (the first bracket) and the $GL(d, \mathbb{R})$ algebra (the third bracket) as immediately identifiable and geometrically significant subalgebras.  

\m 

\n\n{\bf Remark 4} In the 1-$d$ case, 
\be 
G  \es  x \, \d_x 
   \es      D                      \m , 
\label{G-Def-1-d}
\ee
\n{\bf Structure 2} The algebra is now   
\be 
\mbox{\bf [} P \mbox{\bf ,} \, P \mbox{\bf ]} = 0      \m ,
\label{P-P-1-d-P}
\ee 
\be 
\mbox{\bf [} P \mbox{\bf ,} \, D \mbox{\bf ]} = P      \m ,
\label{P-D-1-d-P}
\ee
\be   
\mbox{\bf [} D \mbox{\bf ,} \, D \mbox{\bf ]} = 0      \m , 
\label{D-D-1-d-P}
\ee
\be 
\mbox{\bf [} Q \mbox{\bf ,} \, Q \mbox{\bf ]} = 0      \m ,
\label{Q-Q-1-d-P}
\ee 
\be 
\mbox{\bf [} Q \mbox{\bf ,} \, D \mbox{\bf ]} = - Q    \m ,
\label{Q-D-1-d-P}
\ee
\be   
\mbox{\bf [} P \mbox{\bf ,} \, Q \mbox{\bf ]} = D      \m . 
\label{P-Q-1-d-P}
\ee
The sixth bracket now encodes the 1-way mutual integrability 
\be 
(P, Q) \, \, \Thomas \, \, D                           \m .  
\label{PQ->D}
\ee
\n This 1$d$ projective algebra $Proj(1) = PGL(1, \mathbb{R})$ moreover supports the subalgebras of Fig \ref{Bra-Proj}.a).   

\m 

\n\n{\bf Structure 3} Finding further geometrically significant algebras benefits from making one or both of the symmetric--antisymmetric and trace--tracefree splits of ${G^A}_B$.  
The double split gives a 5-block rendition of the projective algebra $Proj(d)$, as follows. 
\be 
\mbox{\bf [}  P_A \mbox{\bf ,} \, P_{B}  \mbox{\bf ]}                =  0                                                                              \m , 
\label{P-P-P}
\ee 
\be 	 
\mbox{\bf [}  P_A \mbox{\bf ,} \,  D  \mbox{\bf ]}                   =  P_A                                                                            \m , 
\label{P-D-P}
\ee 
\be 	 
\mbox{\bf [}  P_A \mbox{\bf ,} \, \Omega_{BC}  \mbox{\bf ]}          =  2 \, \delta_{A[B} P_{C]}                                                       \m , 
\label{P-O-P}
\ee 
\be 	 
\mbox{\bf [}  P_C \mbox{\bf ,} \, \sigma_{AB}  \mbox{\bf ]}         \es  2 \, \delta_{C(A} P_{B)}  \m - \m  \frac{1}{d} \, \delta_{AB} P_C                \m , 
\label{P-S-P}
\ee 
\be 	 
\mbox{\bf [}  D   \mbox{\bf ,} \, D  \mbox{\bf ]}                    =  0                                                                              \m , 
\label{D-D-P}
\ee 
\be 	 
\mbox{\bf [}  D   \mbox{\bf ,} \, \Omega_{AB}  \mbox{\bf ]}          =  0                                                                              \m , 
\label{D-O-P}
\ee 
\be 	 
\mbox{\bf [}  D , \mbox{\bf \,} \sigma_{AB}  \mbox{\bf ]}            =  0                                                                              \m , 
\label{D-S-P}
\ee
\be 	 
\mbox{\bf [}  \Omega_{AB} \mbox{\bf ,} \, \Omega_{CD}  \mbox{\bf ]}  \es  2 \left( \delta_{C[B}\Omega_{A]D} + \delta_{D[B} \Omega_{A]C} \right)        \m , 
\label{O-O-P}
\ee 
\be 	 
\mbox{\bf [}  \sigma_{AB} \mbox{\bf ,} \, \Omega_{CD}  \mbox{\bf ]}  \es  2 \left( \delta_{C(B} \sigma_{A)D} - \delta_{D(B} \sigma_{A)C} \right)       \m , 
\label{O-S-P}
\ee  
\be 	 
\mbox{\bf [}  \sigma_{AB} \mbox{\bf ,}  \, \sigma_{CD} \mbox{\bf ]}  \es  2 \left( \delta_{D(B} \Omega_{A)C} + \delta_{C(B} \Omega_{A)D} \right)       \m ,   
\label{S-S-P}
\ee
\be 	 
\mbox{\bf [}  Q^A     \mbox{\bf ,} \, Q^B      \mbox{\bf ]}          =  0                                                                              \m , 
\label{Q-Q}
\ee
\be 
\mbox{\bf [}  Q^A   \mbox{\bf ,} \, D  \mbox{\bf ]}                  =  - Q^A                                                                          \m ,
\label{Q-D}
\ee  
\be 
\mbox{\bf [} \Omega_{AB} \mbox{\bf ,} \, Q^C \mbox{\bf ]}           \es  2 \, \delta_{E[A}{\delta_{B]}}^C Q^E                                          \m , 
\label{O-Q}
\ee
\be 
\mbox{\bf [} \sigma_{AB} \mbox{\bf ,} \, Q^C \mbox{\bf ]}           \es  2 \, \delta_{E(A}{\delta_{B)}}^C Q^E  \m - \m  \frac{1}{d} \, \delta_{AB}Q^C     \m , 
\label{S-Q}
\ee
\be 
\mbox{\bf [}  P_A     \mbox{\bf ,} \, Q^B  \mbox{\bf ]}              =   {\delta_A}^{(B} {\delta_C}^{D)} 
                                                                        \left(  {\Omega^C}_D + {\sigma^C}_D  \m - \m  \frac{1}{d} \, {\delta^C}_D D \right)        \m . 
\label{P-Q-P}
\ee
\n{\bf Remark 5} The first ten of these brackets are as per the previous section's split affine algebra. 

\m 

\n The eleventh bracket means that the special-projective transformations $\u{Q}$ commute among themselves.  

\m 

\n The twelfth bracket signifies that $\u{Q}$ is a $Dil$-covector. 

\m 

\n The thirteenth bracket means that $\u{Q}$ is a $SO(d)$-vector. 

\m 

\n The fourteenth bracket extends this to $\u{Q}$ being a $SL(d, \mathbb{R})$-covector.  

\m 

\n Finally, the fifteenth bracket encodes the 1-way mutual integrability 
\be 
(\u{P}, \u{Q}) \, \, \Thomas \, \,  ( \u{\u{\Omega}}, \u{\u{\sigma}}, D)                                                                                                   \m .   
\label{PQ->OS}
\ee 
by which if both $\u{P}$ and $\u{Q}$ are kept, no generators can be dropped.  
This is just the symmetric-antisymmetric and trace-tracefree split version of (\ref{PQ->G}).
\m 

\n\n{\bf Remark 6} The lattice of geometrically-significant subalgebras of Fig \ref{Bra-Proj}.b) can be read off this; 
the preceding integrability is crucial in limiting which combinations of generators close consistently.   
%
{            \begin{figure}[!ht]
\centering
\includegraphics[width=0.7\textwidth]{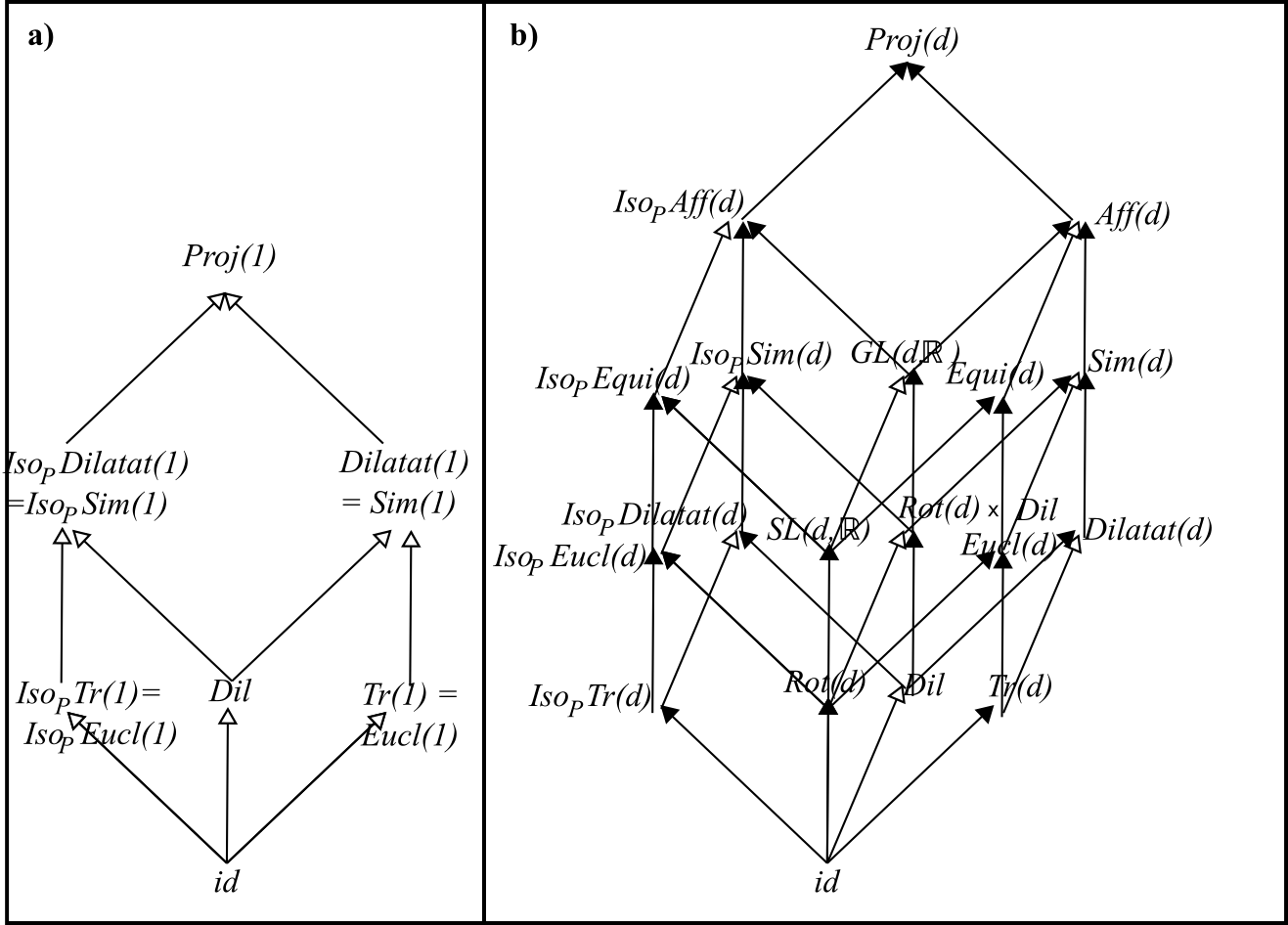}
\caption[Text der im Bilderverzeichnis auftaucht]{        \footnotesize{Geometrically-significant subgroups of $Proj(d)$ a) in 1-$d$ and b) in $\geq 2$-$d$.}}
\label{Bra-Proj} \end{figure}          }

\m 

\n\n{\bf Remark 7} In particular, this admits $P\mbox{-}Iso\mbox{-}Aff(d)$ of $\u{Q}$, $D$, $\u{\u{\Omega}}$ and $\u{\u{\sigma}}$, 
closing according to the 4-block algebra 
\be 	 
\mbox{\bf [}  Q^A     \mbox{\bf ,} \, Q^B      \mbox{\bf ]}          =  0                                                                                     \m , 
\label{Q-Q-I-A}
\ee
\be 
\mbox{\bf [}  Q^A   \mbox{\bf ,} \, D  \mbox{\bf ]}                  =  - Q^A                                                                                 \m ,
\label{Q-D-I-A}
\ee  
\be 
\mbox{\bf [} \omega_{AB} \mbox{\bf ,} \, Q^C \mbox{\bf ]}           \es  2 \, \delta_{E[A}{\delta_{B]}}^C Q^E                                                 \m , 
\label{O-Q-I-A}
\ee
\be 
\mbox{\bf [} \sigma_{AB} \mbox{\bf ,} \, Q^C \mbox{\bf ]}           \es  2 \, \delta_{E(A}{\delta_{B)}}^C Q^E  \m - \m  \frac{1}{d} \, \delta_{AB} Q^C        \m , 
\label{S-Q-I-A}
\ee
\be 	 
\mbox{\bf [}  D   \mbox{\bf ,} \, D  \mbox{\bf ]}                    =  0                                                                                     \m , 
\label{D-D-I-A}
\ee 
\be 	 
\mbox{\bf [}  D   \mbox{\bf ,} \, \Omega_{AB}  \mbox{\bf ]}          =  0                                                                                     \m , 
\label{D-O-I-A}
\ee 
\be 	 
\mbox{\bf [}  D , \mbox{\bf \,} \sigma_{AB}  \mbox{\bf ]}            =  0                                                                                     \m , 
\label{D-S-I-A}
\ee
\be 	 
\mbox{\bf [}  \Omega_{AB} \mbox{\bf ,} \, \Omega_{CD}  \mbox{\bf ]}  \es  2 \left( \delta_{C[B}\Omega_{A]D} + \delta_{D[B} \Omega_{A]C} \right)               \m , 
\label{O-O-I-A}
\ee 
\be 	 
\mbox{\bf [}  \sigma_{AB} \mbox{\bf ,} \, \Omega_{CD}  \mbox{\bf ]}  \es  2 \left( \delta_{C(B} \sigma_{A)D} - \delta_{D(B} \sigma_{A)C} \right)       \m , 
\label{O-S-I-A}
\ee  
\be 	 
\mbox{\bf [}  \sigma_{AB} \mbox{\bf ,}  \, \sigma_{CD} \mbox{\bf ]}  \es  2 \left( \delta_{D(B} \Omega_{A)C} + \delta_{C(B} \Omega_{A)D} \right)              \m ,   
\label{S-S-I-A}
\ee
and its subalgebras,  as per Fig \ref{Bra-Proj}.b). 

\m 

\n\n{\bf Remark 8} These include $P\mbox{-}Iso\mbox{-}Sim(d)$, of $\u{Q}$, $D$ and $\u{\u{\Omega}}$, 
closing according to the 3-block brackets algebra  
\be 	 
\mbox{\bf [}  Q^A     \mbox{\bf ,} \, Q^B      \mbox{\bf ]}          =  0                                                                                     \m , 
\label{Q-Q-I-S}
\ee
\be 
\mbox{\bf [}  Q^A   \mbox{\bf ,} \, D  \mbox{\bf ]}                  =  - Q^A                                                                                 \m ,
\label{Q-D-I-S}
\ee  
\be 
\mbox{\bf [} \omega_{AB} \mbox{\bf ,} \, Q^C \mbox{\bf ]}           \es  2 \, \delta_{E[A}{\delta_{B]}}^C Q^E                                                 \m , 
\label{O-Q-I-S}
\ee
\be 	 
\mbox{\bf [}  D   \mbox{\bf ,} \, D  \mbox{\bf ]}                    =  0                                                                                     \m , 
\label{D-D-I-S}
\ee 
\be 	 
\mbox{\bf [}  D   \mbox{\bf ,} \, \Omega_{AB}  \mbox{\bf ]}          =  0                                                                                     \m , 
\label{D-O-I-S}
\ee 
\be 	 
\mbox{\bf [}  \Omega_{AB} \mbox{\bf ,} \, \Omega_{CD}  \mbox{\bf ]}  \es  2 \left( \delta_{C[B}\Omega_{A]D} + \delta_{D[B} \Omega_{A]C} \right)               \m .  
\label{O-O-I-S}
\ee 
\n{\bf Remark 8} Move (\ref{E-Move}) works here again to establish the isomorphism 
\be 
P\mbox{-}Iso\mbox{-}Sim(d) \m \cong \m  Sim(d) 
                               \es      P\mbox{-}Iso\mbox{-}Tr(d) \rtimes ( Rot(d) \times Dil          ) 
							   \es      \mathbb{R}^d              \rtimes (  SO(d) \times \mathbb{R}_+ ) \m .							   
\ee  
This moreover continues to work alongside 
\be
P^A \rightarrow F^A \mma Q^A \rightarrow - F^A 
\ee 
to establish that also 
\be 
P\mbox{-}Iso\mbox{-}Aff(d)     \es      P\mbox{-}Iso\mbox{-}Tr(d) \rtimes ( Rot(d) \times Dil          ) 
							   \es      \mathbb{R}^d              \rtimes (  SO(d) \times \mathbb{R}_+ )  
						   \m \cong \m                     Aff(d)                                             \m .
\ee

\section{Brackets consistency for a collection of generators}

Let us no longer solve a single GKE for the totality of its GKVs, 
but rather consider which combinations of generators arising from {\it some} GKE are mutually compatible with each other. 

\m 

\n This not only recovers the expected answers -- known notions of geometry -- 
but also gives further answers, alongside specific obstructions to certain combinations of generators being consistent.  

\m  

\n In this approach, $Iso$ versions of geometries have a similar standing to standard geometries. 
They now arise not by picking out subalgebras within solutions of a singe GKE, 
but rather by exploring the other way round: seeing which brackets between generators do not require more generators or are elsewise obstructed. 

\m

\n It is also possible to construct invariants and preserved quantities for $Iso$ geometries \cite{PE-3} by means of our {\sl Seventh} Pillar of Geometry 
(see the Conclusion's Epilogue and \cite{PE-1}) pointing to how some of these geometries arise within other paradigms of Geometry.
See \cite{9-Pillars} for more extensive discussion of this point.

\subsection{The counting argument}\label{Count}

\n This is for finitely-generated automorphism groups. 

\m 

\n We count the number of independent generators, $g$. 

\m 

\n In order to be able to support an $N$-point Shape Theory in dimension $d$, we require that 
\be 
N \, d - g > 0
\ee 
or 
\be 
N \, d - g > 1 \m 
\ee 
if one degree of freedom is to change with respect to another.

\subsection{Special-conformal to special-projective incompatibility} 

\be 
\mbox{\bf [}K_A \mbox{\bf ,} \, Q_B \mbox{\bf ]} = 2 \, \delta_{AB} T
\label{K-Q-Obs}
\ee 
for 
\be
T  :=  ||x||^2 x^C \pa_C   
    =  ||x||^2 D                                                        \m . 
\ee
I.e.\ a 1-way mutual integrability 
\be 
(\u{K}, \u{Q}) \, \, \Thomas \, \,  \u{\u{\u{T}}}                       \m .
\ee 
But $T$ is third-order, by which its own brackets with $\u{K}$ and $\u{Q}$ are fourth order.
Thus, by a straightforward induction, a cascade of arbitrarily high powers of $\u{x}$ forms -- let us denote this by 
\be 
(\u{K}, \u{Q}) \, \, \Thomas \, \,  \u{\u{\u{T}}} \, \, \Thomas \m . \m . \m . \m \infty
\ee 
-- killing off the candidate combination of generators from serving as the automorphism group of a finite Shape Theory.

\subsection{Special-conformal to shear incompatibility} 

\be
\mbox{\bf [} K_A \mbox{\bf ,} \, \sigma_{BC} \mbox{\bf ]}   \es     2 \, \delta_{A(B} K_{C)}   \m + \m \frac{1}{d} \, \delta_{BC} K_A  \m + \m 
                                                                  8 \, \delta_{A(B} Q_{C)} \m - \m \frac{1}{d} \, \delta_{BC} Q_A  \m - \m  4\, R_{A(BC)}   \m ,   
\label{K-S-Obs}
\ee 
for 
\be 
R_{ABC} := x_A x_B \pa_C   \m .  
\ee
Thus, firstly, we have the 1-way mutual integrability 
\be 
(\u{K}, \u{\u{\sigma}}) \, \, \Thomas \, \, \u{Q}    \m ,   
\label{KS->Q}
\ee 
by which special-conformal and shear generators cannot be entertained without also including special projective transformations. 

\m 

\n Secondly, a new $\u{\u{\u{R}}}$ term arises as well: 
\be 
\left(\u{K}, \u{\u{\sigma}}\right) \, \, \Thomas \, \, \left(\u{Q}, \u{\u{\u{R}}}\right)   \m .  
\label{KS->QR}
\ee
Its own brackets result in third-order terms, so a cascade of arbitrarily high powers of $\u{x}$ forms, 
killing off the candidate combination of generators from serving as the automorphism group of a finite Shape Theory:
\be 
\left(\u{K}, \u{\u{\sigma}}\right) \m \Thomas \m \left(\u{Q}, \u{\u{\u{R}}}\right) \m \Thomas  \m . \m . \m . \m  \infty   \m .  
\label{KS->QR->Infty}
\ee

\subsection{$Sim(d)$ extension alternative} 

\n Extending $Sim(d)$ thus involves an {\bf alternative}:

\m 

\n{\bf either} add shear and special-projective generators, 

\m 

\n{\bf or} add special-conformal generators.

\subsection{1-$d$ counterpart} 

\n It is also instructive to contemplate the 1-$d$ situation: here $K$ and $Q$ are not only compatible but also coincide up to sign: 
\be 
K_A  \es  ||x||^2\d_x - 2 \, x^2\d_x 
     \es  - x^2\d_x  
	  =   - Q_A
\ee 
Thus, exceptionally, 
\be  
\mbox{\bf [} K \mbox{\bf ,} \, Q \mbox{\bf ]}  \es    \mbox{\bf [} - Q \mbox{\bf ,} \, Q \mbox{\bf ]} 
                                               \es  - \mbox{\bf [}   Q \mbox{\bf ,} \, Q \mbox{\bf ]} 
											   \es  - 0 
											    =     0                                                 \m , 
												\label{K-Q-1-d}
\ee 
so termination occurs, but $K$ was never independent of $Q$ in the first place, 
so this just returns the 1-$d$ algebra 
\be 
P\mbox{-}Iso\mbox{-}Tr(1) \m \cong \m  C\mbox{-}Iso\mbox{-}Tr(1)  
                          \m \cong \m                      Tr(1)
\ee 
\n In 1-$d$, there are moreover no shears, so our brackets-compatibility analysis is complete.

\section{Brackets Consistency from Polynomial ans\"{a}tze}

We now cease to make even piecemeal use of GKEs to extract generators.
Instead, we posit polynomial ans\"{a}tze, 
and rely on just Lie brackets consistency to extract geometrical automorphism groups that are suitable for supporting Shape Theories.

\subsection{1-$d$ case}

Let us first carry this out in this simpler case.  

\m 

\n{\bf Preliminary Lemma} i) Assuming $\d_x$ returns $Tr(1) = Eucl(1)$. 

\m 

\n ii) $\langle x \, \d_x \rangle$ returns $Dil(1)$.  

\m

\n iii) $\langle \d_x, \, x \, \d_x \rangle$ returns $Dilatat(1) = Sim(1)$. 

\m

\n iv) $x^2\d_x$ returns $P\mbox{-}Iso\mbox{-}Tr(1)$.   

\m

\n v) $\langle x \, \d_x, \, x^2 \d_x\rangle$ returns $P\mbox{-}Iso\mbox{-}Dilatat(1)$.   

\m

\n vi) $\langle \d_x, \, x^2 \d_x \rangle$ fails to close alone, reflecting once again the 1-way mutual integrability (\ref{PQ->D}).

\m 

\n vii) $\langle \d_x , \, x \, \d_x, \, x^2 \d_x \rangle $ gives $Proj(1)$. 

\m

\n viii) Trying to extend using any cubic or higher terms 
\be 
x^{n} \d_x \mma n \geq 3 
\ee 
produces a cascade.  

\m

\n{\bf Remark 1} Geometer Heinrich Guggenheimer pre-empted results vii) and viii) in his 1963 book \cite{G63}.  
He also gave ii) and iv)'s generators. 
We go further by considering the two of these together. 
By associating $P\mbox{-}Iso$ gometries in their own right to iv) and v). 
And by reinforcing the cascade with a counting argument precluding $N$-point Shape Theory 
(as well as $N$-point invariants: a notion that Guggenheimer -- and \'{E}lie Cartan \cite{Cartan55} -- already possessed).  

\m 

\n{\bf Remark 2} I since found that vii) and viii) can furthermore be readily inferred from Chapters 3 and 4 of Lie's 1880 treatise \cite{Lie80}.

\subsection{Higher-$d$ case}

\n We also go further than Guggenheimer by considering higher-dimensional ans\"{a}tze.
The general (bosonic vectorial) quadratic generator in $\geq 2$-d is the following 2-parameter family ansatz
\be 
Q^{\st\sr\si\sa\sll}_{\mu, \nu \, A}  \:=  \mu ||x||^2\pa_A + \nu \, x_A (\u{x} \peq \u{\nabla} )                             \m .  
\ee 
This follows from considering the general fourth-order isotropic tensor contracted into a symmetric object $x^Ax^B$.  

\m 

\n{\bf Theorem} For $d \geq 2$, $\u{Q}^{\st\sr\si\sa\sll}_{\mu, \nu}$ self-closes only if either 
\be 
\mu = 0 
\label{P}
\ee 
or 
\be 
\nu = - 2 \, \mu                                                                                                             \m . 
\label{C}
\ee 
{\bf Remark 1} The first of these is a recovery of the special-projective generator $\u{Q}$, 
     whereas the second is a recovery of the special-conformal generator            $\u{K}$. 

\m 

\n{\u{Proof}} 
\be 
\mbox{\bf [} Q^{\st\sr\si\sa\sll}_{\mu, \nu \, A}  \mbox{\bf ,} \,  Q^{\st\sr\si\sa\sll}_{\mu, \nu \, B}  \mbox{\bf ]}  
                                                        \es    2 \, \mu \, (2 \mu + \nu) ||x||^2 x_{[A}\pa_{B]}  
                                                               \=:    \, \mu \, (2 \mu + \nu) \bigtau_{AB}                        \m   
\ee 
for 
\be 
\bigtau_{AB} := ||x||^2 2 \, x_{(A)} \pa_{B)} = ||x||^2 \Omega_{AB}                                                               \m . 
\ee 
This gives rise to the following cases. 

\m 

\n Case 1) This bracket could vanish strongly, i.e.\ by setting the numerical prefactor to zero. 
This numerical prefactor moreover has two factors that are each capable of being zero, so this case contains two subcases; (\ref{P}) and (\ref{C}) are the corresponding roots.  

\m 

\n Case 2) Elsewise, a third-order term $\u{\u{\bigtau}}$ is produced, for which brackets consistency produces a cascade, 
\be 
\u{Q}^{\st\sr\si\sa\sll} \, \, \Thomas \, \, \u{\u{\bigtau}} \, \, \Thomas \m . \m . \m . \m \infty                                                                               \m .
\ee 
Our counting argument precludes this case from supporting a Shape Theory, so we are done.   $\Box$ 

\m
	 
\n{\bf Corollary 1} Each of $\u{Q}$ and $\u{K}$ is self-consistent, returning $P\mbox{-}Iso\mbox{-}Tr(d)$ 
                                                                          and $C\mbox{-}Iso\mbox{-}Tr(d)$ respectively.

\m 

\n The following supporting Lemma enables a number of further Corollaries. 

\m 

\n{\bf Supporting Lemma} i) The general zeroth order generator $P_A := \pa_A$ is self-consistent according to (\ref{P-P}), returning $Tr(d)$.  

\m 

\n ii) The general scalar and 2-tensor linear ans\"{a}tze      $D := x^A\pa_A$ 
                                                      and $G_{AB} := x_A\pa_B$ are each self-consistent according to (\ref{D-D}) and (\ref{G-G}), 
returning $Dil$  and $GL(d, \mathbb{R})$ respectively.  

\m 

\n In fact, for $d \geq 2$, ansatz $D$ is redundant since this the trace part of $G_{AB}$: 
\be 
{G^A}_A = x^A \pa_A = D \m ,  
\ee 
whereas $d = 1$ exhibits the coincidence (\ref{G-Def-1-d}).
\n iii) These zeroth- and first-order ans\"{a}tze $\u{P}$ and $\u{\u{G}}$ are additionally mutually consistent as per (\ref{P-P}, \ref{G-G}), 
returning the affine algebra $Aff(d)$.  

\m 

\n The preceding coincidence and 

\be 
\mbox{1-$d$'s lack of room for any antisymmetry } \m \Omega = 0 
\label{No-Anti}
\ee 
mean that $Aff(1)$ reduces to $Sim(1)$ and further to $Dilatat(1)$.

\m

\n{\bf Corollary 2} The $\u{Q}$ emerging from the Theorem's first strongly vanishing root 
is mutually compatible with both the zeroth- and first-order ans\"{a}tze $P$ and $G_{AB}$, 
by which $Proj(d) = PGL(d + 1, \mathbb{R})$'s algebra emerges.

\m 

\n $\u{Q}$ can be combined with just $\u{\u{G}}$,                             giving $P\mbox{-}Iso\mbox{-}Aff(d)$, 
                           with just its trace part $D$,                   returning $P\mbox{-}Iso\mbox{-}Dilatat(d)$, 
                           with just its antisymmetric part $\u{\u{\Omega}}$, producing $P\mbox{-}Iso\mbox{-}Eucl(d)$,
                        or with both of these,                              yielding $P\mbox{-}Iso\mbox{-}Sim(d)$.  
It can also be combined with just $\u{\u{G}}$'s tracefree part $\u{\u{\Sigma}}$. 

\m 

\n In 1-$d$,  by (\ref{G-Def-1-d}, \ref{No-Anti}), the first, second and fourth of these conf\n late to $P\mbox{-}Iso\mbox{-}Dilatat(1)$, 
whereas the third and fifth conflate to $P\mbox{-}Iso\mbox{-}Tr(1)$.  

\m 

\n $\u{Q}$ cannot however be combined with the tracefree symmetric part $\u{\u{\sigma}}$ in the absense of the antisymmetric part $\u{\u{\Omega}}$, 
by the 1-way integrability (\ref{S->O}).  

\m 

\n Finally, $\u{Q}$ cannot be combined with the zeroth-order ansatz $\u{P}$ in the absense of the linear ansatz $\u{\u{G}}$ 
or of any irreducible part thereof, by the 1-way integrability (\ref{PQ->G}).  

\m 

\n{\bf Corollary 3} The $\u{K}$ emerging from the previous Theorem's second strongly vanishing root 
is mutually compatible with the zeroth-order ansatz $\u{P}$ alongside the first-order ansatz's trace and antisymmetric parts $D$ and $\u{\u{\Omega}}$. 
In this way, the relation (\ref{Conf(d)}) emerges for $d \geq 3$, and the $SO(3, 1)$ subalgebra of the infinite $Conf(2)$ algebra emerges for $d = 2$.  

\m 

\n $\u{K}$ cannot however be combined with the symmetric part of the linear ansatz by the cascade-sourcing obstruction (\ref{K-S-Obs}).  

\m 

\n $\u{K}$ can be combined with just $D$,    giving $C\mbox{-}Iso\mbox{-}Dilatat(d)$, 
              with just $\u{\u{\Omega}}$, returning $C\mbox{-}Iso\mbox{-}Eucl(d)$, 
                   or with both of these,  yielding $C\mbox{-}Iso\mbox{-}Sim(d)$.    

\m 

\n{\bf Remark 2} Via some parts of the Supporting Lemma and the first part of each of Corollaries 2, and 3, 
our `Brackets Consistency' Pillar of Geometry {\sl derives} both Projective and Conformal Geometry in flat space.  
This is moreover a {\sl conceptually new type} of derivation from previous ones in the literature, 
by which a new foundational paradigm for each of Projective and Conformal Geometry in flat space.  
These moreover arise side by side as the two roots of an algebraic quadratic equation that emerges as the right-hand-side of a Lie bracket.

\m 

\n{\bf Remark 3} This method does not pick out the infinite-$d$ extension of the conformal group in 2-$d$. 

\m

\n From the perspective of $N$-point Shape Theory \cite{Kendall84, Kendall89, Small, Kendall, Bhatta, PE16, S-I, S-II, S-III, Minimal-N},  
however, this does not matter, since this case's infiniteness precludes it from supporting such a Shape Theory [by Sec \ref{Count}'s counting argument].

\m 

\n Another qualification, independent of Shape Theory, 
is that our methodology just returns the {\sl finitely generated} geometrical automorphism groups. 
It is moreover logically possible for some of the infinite cascades excluded by our method 
to have significance as infinitely-generated automorphism groups, and this then so happens to be realized in the case of $Conf(d)$.

\m 

\n Of course, using `Killing's' Fifth Pillar, this case is detected by the flat CKE collapsing in 2-$d$ to the Cauchy--Riemann equations \cite{AMP}.  
The loss of this immediate deduction if one uses instead the `Brackets Closure' Sixth Pillar is a first example of a price to pay for using 
foundations that make less structural assumptions.
As argued above, this is significant if one's remit is Foundations of Geometry, but not if it is Shape Theory.  

\m  

\n{\bf Remark 4} Thus both `top geometries' in flat space -- conformal and projective -- arise as the 2 roots of a single algebraic equation 
for the strong vanishing of the self-bracket of the general quadratic ansatz generator.

\m 

\n{\bf Remark 5} In 1-$d$, this result fails because antisymmetric 2-forms are not supported. 
It fails moreover ab initio since there is only one rank-4 isotropic tensor in 1-$d$: the constant, 
by which our working collapses to finding the bracket of a scalar with itself, which is of course trivially zero. 
It is now clear why the theorem excludes 1-$d$, and how earlier workings in the current paper recover the outcome of the simple special case $d = 1$.   

\m 

\n{\bf Remark 6} Lie's \cite{Lie80} own systematic classification of Lie algebras only went as high as 2-$d$.  
See \cite{Olver, Olver2} for contemporary review of Lie Theory including its Flow PDEs and Invariants aspects.

\section{Conclusion}

\subsection{Commentary}

We considered a layer-by-layer removal of generalized Killing assumptions leading to Section 5's `Brackets consistency' Sixth Pillar of Geometry. 
This is a clear enough development to not require a summary here, 
this Conclusion serving instead to follow up Section 5's `Brackets Consistency' Sixth Pillar of Geometry with some far-reaching comments. 

\m 

\n \cite{Higher-Lie, XIV} has more to say about such strongly-vanishing numerical prefactors with multiple physically and/or geometrically significant factors 
pervading both Physics and Geometry once considered within the Brackets Consistency Paradigm. 
This is significant in replacing {\sl considerable insights} with straightforward algorithmic calculations that {\sl anybody} could do. 
E.g. both the fork between universal relativity being Galilean or Lorentzian, and the dynamical structure of GR,  
drop out \cite{RWR, San, OM02, OM03, AM13, ABook}  from such strong vanishings in the context of physical constraints brackets.
The suggestion is then that working extensively enough within the Brackets Consistency Paradigm 
would lead algorithmically to results which would elsewise require significant insights, 
including in fields that have not yet been blessed with insightful enough theoreticians to proceed otherwise...   

\m 

\n Within the Theoretical Physics context, this could be paraphrased as 
`Dirac established a paradigm \cite{Dirac, Sni, S82, HT92} 
by which previous classical theoretical insights could be recovered \cite{RWR, OM02, OM03, Phan, AM13, ABook} merely by solving an algorithm' (!) 
The current article, moreover, provides a first major example of how the {\sl brackets consistency} 
aspect of Dirac's Algorithm by itself has enough power to produce results. 
This application can furthermore be abstracted well outside the remit of Dirac's Algorithm, 
i.e.\ not only for  a classical Theoretical Physics phase space with constraints 
and thus a constraint algebraic structure thereupon, but also wherever there is just a brackets algebra.   
Restricting attention to Lie brackets in Geometry, Brackets Consistency can in this way be viewed as a new Pillar of Geometry. 
This Pillar's technical tools have been known since Cartan in the 30s through to the 50s \cite{Cartan55} or even to Lie \cite{Lie} in the 1890s, 
but appreciation that these constitute a Pillar had to await the above recoveries and subsequent abstraction.  
This Pillar gives a conceptually new derivation for $d \geq 3$ of the conformal versus projective alternative as regards flat-space geometry corresponding 
to a top automorphism group.   

\m 

\n This is the final way in which we go further than Guggenheimer, noting that his book \cite{G63} dates from 1963 whereas Dirac's one \cite{Dirac} 
expounding his Algorithm is from 1964 (and in a different field of study besides).  
We subject this to a considerable amount of comparison, Socratic thinking and cross-fertilization in the upcoming Article `9 Pillars of Geometry' \cite{9-Pillars}. 
We also compare this approach further with its more elaborate phase space cousin the Dirac Algorithm for consistency of constraint algebraic structures 
in another recent Article \cite{XIV}.  

\m 

\n Some other recent literature on applications of the Dirac Algorithm includes \cite{PG13, B15, B16, BOP17}. 
We however distinguish between treating individual models or theories on the one hand, and passing whole candidate families of theories through the Dirac Algorithm. 
We also distinguish between demanding solely consistency and placing further demands, such as bracket right-hand-side simplifications \cite{BK94, BM96} 
or the presupposition of spacetime structure \cite{HKT, T80}.  
This last case is, however, clearly not a Spacetime Reconstruction.  
We also generalized from the Dirac Algorithm to a substantially improved form of Lie Algorithm in \cite{Higher-Lie, XIV}, 
as well as pointing to deformation, rigidity and subsequent cohomological underpinnings \cite{G64, NR66}.
Finally, this Article's approach has been shown to extends to further conceptions of brackets, 
such as Nambu brackets \cite{Nambu} or Nijenhuis/Gerstenhaber type brackets, as are suitable e.g.\ for quantum operator algebras.

\subsection{Epilogue}

Some more Pillars of Geometry are as follows (nor is this meant to be an exhaustive list; \cite{XIV, 9-Pillars} further compare these).  

\m 

\n{\bf The Seventh Pillar of Geometry} is \cite{PE-1, PE-2, PE-3} arriving at preserved quantities -- smooth functions of invariants -- 
end of the Erlangen Pillar by solving a different PDE to Killing's: the preserved equation. 

\m 

\n[This is moreover the restriction to configuration space \cite{DO-1, XIV} 
of Expression in terms of Observables aspect of Background Independence \cite{K92, I93, APoT2, ABeables, ABook}.]

\m 

\n{\bf The Eighth Pillar of Geometry} is the corresponding differential (rather than PDE-integrating) theory of invariants due to Cartan \cite{Cartan55}.

\m 

\n{\bf The Ninth Pillar of Geometry} is Supersymmetry as a first principle for Geometry. 
In its simplest manifestation, supersymmetry can be envisaged as the square root of translation, and represents trading in ordinary space for Grassmann space. 
This adds dimensions, and of a qualitatively different kind, starting with a different arithmetic being introduced.  

\m 

\n(This additionally has parallels with Supersymmetry being a first principle in Background Independence, 
on a very comparable footing to Temporal Relationalism and Configurational Relationalism \cite{ABeables, AMech, ABook}.)  

\m

\n This requires `super' upgrades of many Pillars, in particular of the following.  

\m 

\n{\bf Super-Pillar 2} is now not      Algebra from Cartesian    Geometry                   through to Linear      Algebra, 
                       but rather Superalgebra from Grassmannian Geometry \cite{DeWitt84, V04} through to Superlinear Algebra \cite{V04}.

\m 

\n{\bf Super-Pillar 3} is based on super-transformation supergroups, 
forming super-geometrically-significant super-automorphism supergroups with an associated theory of super-invariants and super-geometric preserved quantities. 
  
\m 

\n[Aside from this `Super-Erlangen variant, the Erlagen approach may be taken to apply rather more widely than this: 
one can argue for it to be a {\sl categorical} approach to all of mathematics.
Namely, the morphism-centric perspective indeed to be found in many a modern exposition of whichever branch of Pure Mathematics.]

\m 

\n{\bf Super-Pillar 5} Granted first the existence of the Killing spinor equation \cite{F00} 
and (in particular conformal, affine, projective) generalizations thereof, 
we have super-Killing Zeroth Principles on which the supergroups of super-Erlangen rest 
(with some benefits from supersymmetry enhancing integrability).  

\m 

\n{\bf Super-Pillar 6} Super-Lie brackets support a corresponding notion of Super-brackets Closure: 
an assessment of which combinations of super-geometric generators are consistent.

\m 

\n $Super\mbox{-}Sim(d)$ extension is itself a third alternative to $Proj(d)$ and $Conf(d)$; 
it is based now not on adding quadratic generators but on further generalizing linear generators in the sense of Grassmann.
Moreover this one combines with each of conformal and projective separately, in each case producing a necessarily 
$N = 2$ supersymmetries: both $\u{K}$ and $\u{Q}$ are `sufficiently translation-like' supply a second supersymmetry generator.  
So we have two competing top supergroups: $Super\mbox{-}Conf(d)$ and $Super\mbox{-}Proj(d)$. 
Each of these moreover substantially exceed the sums of their parts, rendering them even more veritable super-top geometries.

\m 

\n Supersymmetry offers another way into linear theories, by having a linear ansatz that now includes linearity in the new Grassmann variable. 
This accounts for their being missed out in both Classical Geometry and Section 5's outline of Pillar 6. 

\m 

\n Demanding subalgebras that are themselves supersymmetric is a notable subcase within Super-Pillar 6. 

\m  

\n{\bf Super-Pillar 7} is to solve super-preserved equations to obtain super-preserved quantities: 
smooth functions of $N$-point super-invariants.
This is a Zeroth Principle for the super-invariants end of super-Erlangen.  

\m 

\n{\bf Super-Pillar 8} is a super-Cartan differential approach to super-invariants.  

\m 

\n This Epilogue sets the scene for the Author's upcoming supersymmetric counterpart of \cite{PE-1, DO-1} and the current Article, 
with \cite{XIV, 9-Pillars} providing some of the next steps.   

\m 

\n{\bf Acknowledgments} I thank Chris Isham, Niall O'Murchadha, Don Page, S. Brahma, and an anonymous socratic thinker for previous discussions, 
as well as Malcolm MacCallum, Jeremy Butterfield, Reza Tavakol and Enrique Alvarez for support with my career.  


\begin{appendices}

\section{Block splits of Lie algebras}

Let $\Frg$ be a Lie algebra \cite{Gilmore, Serre}, with generators $g_A$.   
\beq
\mbox{\bf [} g_{A} \mbox{\bf ,} \, g_{B} \mbox{\bf ]} = {G^{C}}_{AB} \, g_{C}   \m ,
\label{Str-Const}
\eeq 
for ${G^{C}}_{AB}$ the corresponding {\it structure constants}.

\m 

\n Suppose that a hypothesis is made about some subset $\Frk$ of generators $k_{K}$ being significant.    
Let us denote the set of the rest of the generators $h_{H}$ by $\Frh$.   
Such a hypothesis can amount to seeing whether a model in hand implements some candidate principle. 
This could for instance be one or more of an aesthetic principle, 
                                           a philosophical principle, 
										   a strategic principle, 
										or a program-defining principle. 
The Problem of Time literature contains an example of all four at once \cite{K93, AObs3, ABook}: 
whether the linear and quadratic constraints are to be treated qualitatively differently from each other.  

\m 

\n One next needs moreover to check whether the algebraic structure at hand actually complies with this assignation of significance.    
If a counter-example to the hypothesis holding can thus be found, the hypothesis is either not universal or requires qualification. 
Further considerations then include whether the hypothesis is generically true, 
and whether it is true in elsewise significant, indispensible cases. 
This illustrates how computing brackets algebras has the capacity to dismiss some candidate hypotheses about 
some subset of generators being qualitatively different from the remainder.

\m

\n The general algebraic structure for a 2-block split is of the form 
\beq
\mbox{\bf [} k_K \mbox{\bf ,} \,  k_{K^{\prime}} \mbox{\bf ]}  \es  {A^{K^{\prime\prime}}}_{K K^{\prime}} k_{K^{\prime\prime}}   \m + \m 
                                                                      {B^H}_{K K^{\prime}}                  h_{H}                     \m ,
\label{Lie-Split-1}
\eeq 
\beq
\mbox{\bf [} k_K \mbox{\bf ,} \,  h_H          \mbox{\bf ]}    \es  {C^{K^{\prime}}}_{K H} k_{K^{\prime}}                        \m + \m 
                                                                      {D^{H^{\prime}}}_{K H} h_{H^{\prime}}                           \m ,
\label{Lie-Split-2}
\eeq 
\beq
\mbox{\bf [} h_H \mbox{\bf ,} \,  h_{H^{\prime}} \mbox{\bf ]}  \es  {E^K}_{H H^{\prime}}                  k_K                    \m + \m
                                                                      {F^{H^{\prime\prime}}}_{H H^{\prime}} h_{H^{\prime\prime}}      \m .
\label{Lie-Split-3}
\eeq
This is entertained e.g.\ in Gilmore's book \cite{Gilmore}.  
 
\m  

\n      $A$ encodes {\it self-interaction} of the $\Frk$ block's generators, 
whereas $F$ encodes that                   of the $\Frh$ block's.   

\m 
 
\n $\Frk$ is a subalgebra if $B = 0$, whereas $\Frh$ is a subalgebra if $E = 0$.   
$B$ and $E$ are thus conceptually `block nonsubalgebraicities': measures of departure from subalgebra status.  

\m 

\n Finally $C$ and $D$ are a type of `interactions between' $\Frh$ and $\Frk$: 
{\it self-contained} interactions, since their cofactors reside within the $\Frk$ and $\Frh$ blocks respectively.  

\m 

\n We next comment on a number of subcases that the main text refers to extensively.

\subsection{Two 1-$d$ blocks}

Let us first suppose that both blocks are generated by individual scalars. 
The most general form now is  
\beq
\mbox{\bf [} k \mbox{\bf ,} \,  k \mbox{\bf ]}   =   0                         \m ,
\label{Lie-Split-11}
\eeq 
\beq
\mbox{\bf [} k \mbox{\bf ,} \,  h \mbox{\bf ]}  \es  c \, k + d \, h           \m ,
\label{Lie-Split-21}
\eeq 
\beq
\mbox{\bf [} h \mbox{\bf ,} \,  h \mbox{\bf ]}   =   0                         \m .
\label{Lie-Split-31}
\eeq
This has 3 distinguished cases: 

\m 

\n I) $c = 0 = d$ is a {\it direct product}, denoted by 
\be 
\Frg = \Frk  \times \Frh  \m . 
\ee 
\n II) $c = 0, d \neq 0$ is the {\it semidirect product}, denoted by 
\be 
\Frg = \Frk  \rtimes \Frh  \m . 
\ee 
Note that this is {\sl directed}, as the $\Frk$ block is {\sl normal} $\lhd$ in $\Frg$ whereas the $\Frh$ block need not be.  
Such products are of the form \cite{Cohn}
\be 
(k_1, h_1) \m \circ \m  (k_2, h_2) = (k_1 \, * \, \varphi_{k_1}(n_2), h_1 \Box h_2)
\ee 
for $\langle \Frk, \, * \rangle  \lhd (\lFrg, \circ)$, $\langle \Frh, \Box \rangle \leq (\lFrg, \circ)$ 
and $\varphi:\FrH \rightarrow Aut(\FrN)$ \index{automorphism} a group homomorphism. 
By this last structure, being a semidirect product is not by itself a unique specification; that would require giving $\varphi$ as well.

\m 

\n III) $c \neq 0, d = 0$ is the semidirect product the other way around, i.e.\ 
\be 
\Frg = \Frh  \rtimes \Frk  \m . 
\ee

\subsection{A 1-$d$ block and a $\geq 2$-$d$ block}

Next, suppose one block is a scalar and the other is a non-scalar tensor. 
\beq
\mbox{\bf [} k \mbox{\bf ,} \,  k \mbox{\bf ]}    = 0                                                                           \m ,
\label{Lie-Split-12}
\eeq 
\beq
\mbox{\bf [} k \mbox{\bf ,} \,  h_H \mbox{\bf ]}  \es  {C^{K^{\prime}}}_{H} k_{K^{\prime}} \m + \m  {D^{H^{\prime}}}_H h_{H^{\prime}}  \m ,
\label{Lie-Split-22}
\eeq 
\beq
\mbox{\bf [} h_H \mbox{\bf ,} \,  h_{H^{\prime}} \mbox{\bf ]} \es   E_{H H^{\prime}}                     k                    \m + \m            
                                                                   {F^{H^{\prime\prime}}}_{H H^{\prime}} h_{H^{\prime\prime}}     \m .
\label{Lie-Split-32}
\eeq 
This has as distinguished cases, firstly, the previous section's trio with an enlarged list of conditions for each. 

\m 

\n I) $C = 0 = D \mma E = 0 = F$ is a direct product           $\Frh  \times \Frk$.

\m

\n II) $C = 0, D \neq 0 \mma E = 0$ is a semidirect product    $\Frk  \rtimes \Frh$.

\m 

\n III) $c \neq 0, d = 0 \mma E = 0$ is the semidirect product $\Frh  \rtimes \Frk$.   

\m 

\n Secondly, a new possibility is supported by one block being $\geq 2$-$d$. 

\m  

\n IV) $E \neq 0$ we have a {\it 1-way integrability} \cite{ABook}, denoted by 
\be 
\Frh \Thomas \Frk                                                                                                                       \m .  
\ee 
In this case, $\Frk$ is not a subalgebra: attempting to close it leads to discovering the necessity of including $k_{\sfk}$ as well.

\subsection{Two $\geq 2$-$d$ blocks}

Let us finally consider the furtherly general case, in which both blocks are generated by non-scalar tensors \cite{ABook}. 
This has as distinguished cases, firstly, the previous section's quartet with an enlarged list of conditions for each. 

\mbox{ } 

\ni I)    If $B = C = D = E = 0$, we have a direct product                 $\Frh  \times \Frk$.   

\m

\ni II)   If $B = C = E = 0$ but $D \neq 0$, we have a semidirect product  $\Frk  \rtimes \Frh$.   

\m 

\ni III)  If $B = D = E$ but $C \neq 0$, we have a semidirect product      $\Frh  \rtimes \Frk$.  

\m

\ni IV)   If $E \neq 0$ but $B = 0$, we have the one-way integrability     $\Frk \, \Thomas \, \Frh$.

\m 

\n A simple and well-known example of IV) occurs in splitting the Lorentz group's generators up into blocks of 3 boosts and of 3 rotations. 
Here the rotations close as a subalgebra, whereas the boosts imply that the rotations must be included as well.  
This is the group-theoretic underpinning \cite{Gilmore, Weinberg-1} of {\it Thomas precession}, 
furnishing the alternative pedagogically useful name {\it Thomas integrability} \cite{ABook} for 1-way integrability. 

\m 

\n Secondly, two new possibilities are supported by both blocks being $\geq 2$-$d$. 

\m 

\ni V)    If $E = 0$ but $B \neq 0$, we have one-way integrability in the other direction: 
\be 
\Frh \, \Thomas \, \Frk  \m . 
\ee 
\ni VI)  If both $B \neq 0$ and $E \neq 0$, we have {\it two-way integrability}, denoted 
\be 
\Frk \, \TwoWay \, \Frh \m .
\ee 
Now neither $\Frk$ nor $\Frh$ are subalgebras; rather, each's generators implies (some of) the other's.  
In this case, any incipient desires for $\Frk$ to play a significant role by itself are almost certainly dashed by the 
mathematics of the ensuing algebraic structure.

\m

\ni Note that, sequentially, VI), then IV) and V) and then I)-III) represents a progression with sharp decrease in mathematical diversity.  

\m

\n Similar analysis applies to splits into 3 or more blocks (the current Article goes as high as five, and its super-companion paper as high as seven).

\subsection{Integrabilities, obstructions, infinite cascades and strong vanishing}

Start with one block $\Frh$. 
Suppose that 
\be
\mbox{\bf [} h_A \mbox{\bf ,} \,  h_{A^{\prime}} \mbox{\bf ]} = {B^B}_{AA^{\prime}} \, k_{B}
\ee 
(which is possible for $\mbox{dim}\, \Frh \geq 2$).  
Then the $\Frh$ block self-implies a second block $\Frk$. 

\m 

\n VII) It is also possible to have 
\be 
\mbox{\bf [} h_A \mbox{\bf ,} \,  h_{A^{\prime}} \mbox{\bf ]} = \Theta_{AA^{\prime}} 
\ee 
for $\u{\u{\Theta}}$ a {\it hard self-obstruction}, or indeed 
\be
\mbox{\bf [} h_A \mbox{\bf ,} \,  h_{A^{\prime}} \mbox{\bf ]}  \es  {A^{A^{\prime\prime}}}_{AA^{\prime}} \, h_{A^{\prime\prime}} \m + \m  
                                                                    {B^B}_{AA^{\prime}} \, k_B                                   \m + \m \Theta_{AA^{\prime}}  \m .
\ee 
\n VIII) Next suppose we have two blocks $\Frh$ and $\Frk$, and 
\be 
\mbox{\bf [} h_A \mbox{\bf ,} \, k_B \mbox{\bf ]}       \es  {G^{C}}_{AB}          \, j_C                            \m .
\ee    
We term this a {\it mutual integrability}. 

\m 

\n IX) It is also possible to have 
\be 
\mbox{\bf [} h_A \mbox{\bf ,} \,  k_B \mbox{\bf ]} = \Theta_{AB}   \m , 
\ee 
for $\u{\u{\Theta}}$ now {\it hard mutual obstruction}, or indeed

\be
\mbox{\bf [} h_A \mbox{\bf ,} \, k_B \mbox{\bf ]}       \es   {A^{A^{\prime}}}_{AB} \, h_{A^{\prime}}  \m + \m  
                                                              {B^{B^{\prime}}}_{AB} \, k_{B^{\prime}}  \m + \m 
   												   		      {J^{C}}_{AB}          \, j_C             \m + \m  \Theta_{AB}   \m .
\ee        
\n Moreover upon discovering $\Frk$, we need to check its self-consistency and its mutual consistency; 
each of these is capable in general of breeding more blocks, or an obstruction. 

\m 

\n X) If a theory has a finite number of degrees of freedom, moreover, an {\it infinite cascade of new blocks} also kills it off.     

\m

\n XI) If cases VII), IX), X)'s extra terms have prefactors which can be adjusted to be zero, 
hard obstructions and infinite cascades can be removed by picking strongly-vanishing coefficients. 

\end{appendices}

\vspace{10in}


\end{document}